%% file: paper.tex
\documentclass[format=acmsmall]{acmart}
\usepackage{float}
\usepackage{subfig}
\usepackage{graphicx}
\usepackage{comment}
\usepackage{amsmath}
\usepackage{amssymb}
\usepackage{mdframed}
\usepackage{xcolor}

\newcommand{\blue}[1]{\textcolor{blue}{#1}}

\begin{document}

\title{Empirical Characterization of Mobility of Multi-Device Internet Users}

\author{Amee Trivedi}
\affiliation{%
  \institution{University of Massachusetts Amherst}
  \country{USA}}
\author{Jeremy Gummeson}
\affiliation{%
  \institution{University of Massachusetts Amherst}
  \country{USA}}
\author{Prashant Shenoy}
\affiliation{%
  \institution{University of Massachusetts Amherst}
  \country{USA}
}

%
%
%
%



\begin{abstract}
\input{abstract}

\end{abstract}

\maketitle

\input{introduction}
\input{background}

\input{dataset}

\input{dev_classification}
\input{multi_device_users}
\input{dev_hierarchical_mobility}

\input{model}

\input{implications}
\input{conclusion}


\bibliographystyle{abbrv}
\bibliography{paper} 

\input{appendix}

\end{document}

%% file: abstract.tex
Understanding the mobility of humans and their devices is a fundamental problem in mobile computing. While there has been much work on empirical analysis of human mobility using mobile device data, prior work has largely assumed devices to be independent and has not considered the implications of modern Internet users owning multiple mobile devices that exhibit correlated mobility patterns. Also, prior work has analyzed mobility at the spatial scale of the underlying mobile dataset and has not analyzed mobility characteristics at different spatial scales and its implications on system design. In this paper, we empirically analyze the mobility of modern Internet users owning multiple devices at multiple spatial scales using a large campus WiFi dataset. First, our results show that mobility of multiple devices belonging to a user needs to be analyzed and modeled as a group, rather than independently, and that there are substantial differences in the correlations exhibited by device trajectories across users that also need to be considered. Second, our analysis shows that the mobility of users shows different characteristics at different spatial scales such as within and across buildings.  Third, we demonstrate the implications of these results by presenting generative models that highlight the importance of considering the spatial scale of mobility as well as multi-device mobility. More broadly, our empirical results point to the need for new modeling research to fully capture the nuances of mobility of modern multi-device users.

%% file: introduction.tex
\section{Introduction}
\label{sec:intro}

Understanding the mobility of users and their devices has become ever more important in the era of the mobile Internet---mobile behavior has broad implications on the design of mobile services, wireless networks, edge computing, and urban infrastructure.  Over the past decade, there has been extensive work on understanding  human mobility at urban scales  \cite{zhao2016urban, xia2018exploring,isaacman2012human,d2017if} and on modeling such mobility \cite{lin2017deep,ijcai2017-430, Ganti:2013:IHM:2493432.2493466, Lee:2006:MST:1132905.1132915, Do:2012:CCM:2370216.2370242, kim2006extracting, DBLP:journals/corr/PappalardoS16, Lin:2012:PIM:2370216.2370274} by using a variety of sources such as  cellular, WiFi, social media check-ins, and vehicular data \cite{hang2018exploring, veloso2011urban, hasan2013understanding,jurdak2015understanding}.
This body of work has largely assumed mobile devices to be independent, an assumption that no longer 
holds in an era of mobile Internet users who own a multitude of devices that exhibit correlated mobility patterns. Further, prior work has analyzed or modeled mobility patterns at a single spatial scale--often that of the underlying dataset---and has not considered the impact of mobility at different spatial scales on system design.


In this paper, we focus on characterizing the mobility of modern mobile
Internet users with a view to answering three questions: 
(1) Since modern users own multiple mobile devices, how correlated or different are the  mobility patterns exhibited by different devices belonging to the same user? 
 (2) How do mobility patterns of devices and users vary across different spatial scales? (3) What are 
 the implications of these findings on problems such as mobility modeling and system design?
We address these research questions by conducting a large-scale study using a campus WiFi dataset from a large university campus comprising 156 buildings
and 5104 WiFi access points; the four month long dataset comprises 
35,699 users and 70,040 devices. We also discuss ethical considerations that arise when conducting
this study ($\S$ \ref{sec:wifi-dataset}).

In conducting our study, this paper makes three main contributions. First, through analysis and modeling, 
we show that it is important to analyze the collection of each user's devices as a group rather than treating them as independent, and also show that correlations in trajectories   within such group can vary significantly across users.  Second, we show that mobility characteristics of devices are different at different spatial scales. Others have  observed, for example, that outdoor mobility differs from mobility of humans indoors. Given our campus  data set, we quantify these differences by examining \emph{micro-scale} mobility  inside buildings (``intra-building scale'') and \emph{macro-scale} mobility across buildings  (``inter-building mobility'').  Finally, we present generative models for multi-device users and mobility at multiple spatial scales and show how these models  yield improvements for mobility problems such as next-location prediction and finding misplaced devices.


\begin{figure}
\centering
\includegraphics[width=0.90\linewidth]{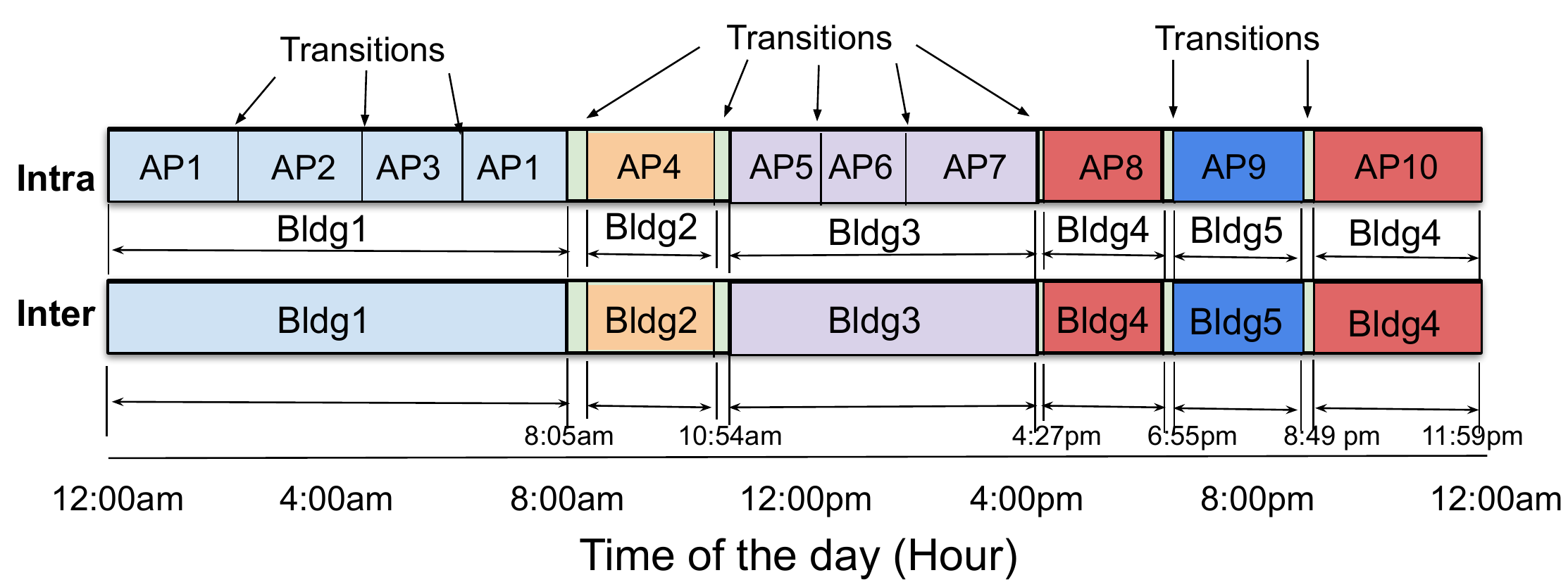}
\caption{Intra and Inter-building Mobility}
\vspace{-0.05in}
\label{fig:hierarchical_view}
\end{figure}

\begin{figure}
\centering
\includegraphics[width=0.90\linewidth]{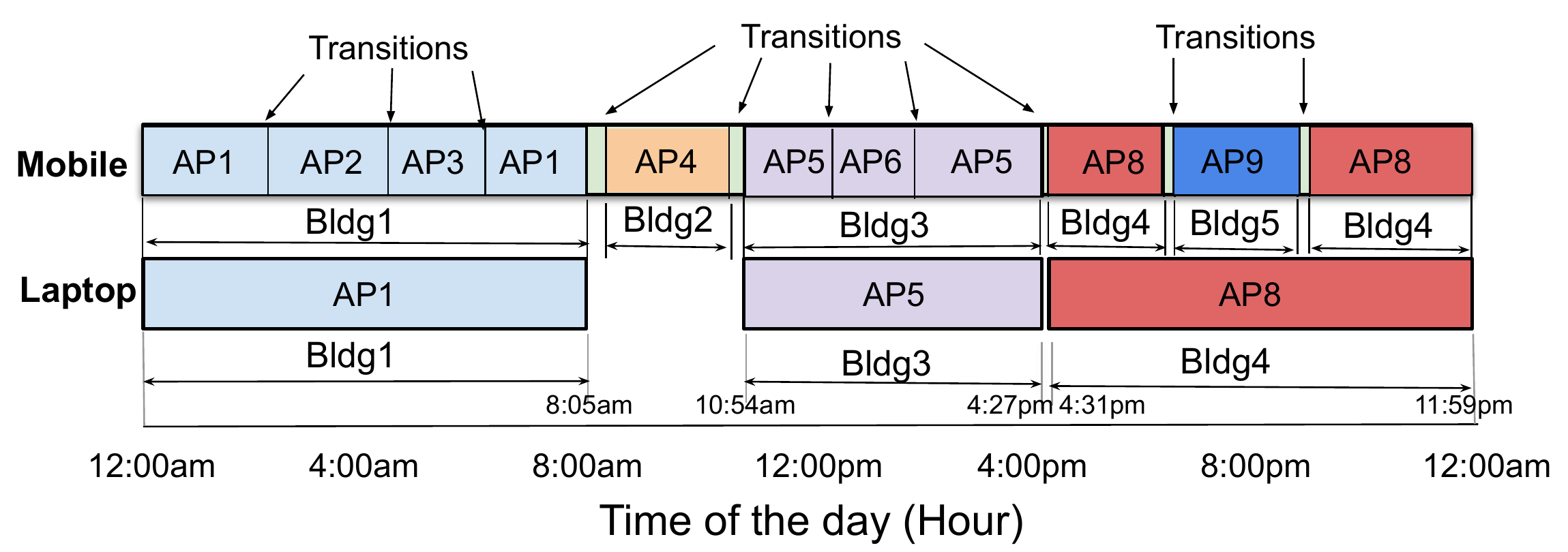}
\caption{Different devices of a user exhibit dissimilar trajectories}
\vspace{-0.05in}
\label{fig:traj_timeline}
\end{figure}


{\noindent \bf Key Results:} Our empirical study reveals many new findings and insights.
First, our analysis reveals 
mobility decreases with increasing spatial scales---we find 8X more mobility at intra-building (micro) scale than inter-building (macro) scale. The opposite is true for time spent at each visited location---inter-building visits have 2X higher stay durations than intra-building ones. 
Second, we find that the type of mobile device has a significant impact on its mobility---phones exhibit  3.5X more mobility in terms of locations visited per day than laptops by virtue of being smaller and more portable, implying that device type matters and should be considered when designing algorithms and systems for a mix of devices.
 Third,  we find that different devices owned by a user 
exhibit moderate to strong correlations in their daily trajectories, but  
 the degree of correlation can vary in significant ways based on the user. 
Finally, we find intra-building and inter-building mobility to be far more frequent, in terms of locations visited per day, than outdoor urban-scale mobility results from prior work, confirming our hypothesis that the spatial scale of mobility should be a key design consideration. 

%% file: background.tex
\section{Background}
\label{sec:background}

\begin{table*}[ht] 
    \begin{center}
     \resizebox{\textwidth}{!}{
        \begin{tabular}{ccccc} \toprule
        Study & Data Source & Spatial Scale & Multi-device users & Use case \\ \midrule
        Tsinghua campus \cite{zhou2017mining} & WiFi, SNMP, Apps & Single, flat & No & Student behavior \\
        SMU campus \cite{Jayarajah15,Jayarajah18} & WiFi, Apps & Single, flat & No & Group/user behavior \\ 
        Dartmouth campus \cite{kotz2005analysis} & Network WLAN& Single, flat & No & Network optimization \\
        Corporate campus \cite{balazinska2003characterizing}& Network WLAN & Single, flat & No & WLAN characterization \\
        Our study & WiFi syslog &  Inter- \& Intra-building & Yes & Modeling \\ \bottomrule
        \end{tabular}}
     \caption{Comparison with Prior Campus-scale Mobility Studies}\label{tab:prior}
 \end{center}
\end{table*}

In this section, we present background on characterizing human mobility using mobile devices.

{\noindent \bf Humans as Nomads}
Human mobile behavior is assumed to be {\em nomadic} in nature, where nomadicity involves traveling to a location, staying at that location for a period of time, followed by travel to a new location, and so on ~\cite{Kleinrock:1996:NAA:274617.274618}.    The process of moving between two successive locations is referred to as a {\em transition}, while the stationary behavior at a location is denoted as a {\em stationary} period. The path of a mobile user over time is referred to as their {\em trajectory}.

Early work in mobile computing focused on characterizing such nomadic behavior as {\em random walks}, where the choice of the next location was random ~\cite{rhee2011levy, Bettstetter:2004:SPR:1027335.1027342, bai2004survey,rhee2011levy}. However subsequent research showed that human activities follow daily and weekly routines, and there are significant spatial and temporal correlations as well as recurring patterns in the locations visited by mobile users ~\cite{gonzalez08, Axhausen2002, Jiang2012, zheng2008understanding, Ganti:2013:IHM:2493432.2493466}. These empirical characterization studies led to new modeling efforts ~\cite{song2010modelling, Ganti:2013:IHM:2493432.2493466, kim2006extracting, Lee:2006:MST:1132905.1132915, Do:2012:CCM:2370216.2370242, kim2006extracting, DBLP:journals/corr/PappalardoS16, Lin:2012:PIM:2370216.2370274} that employed markov models ~\cite{Lee:2006:MST:1132905.1132915, gambs2012next,Lee:2006:MST:1132905.1132915,lee2009slaw,naboulsi2016large} as well as deep learning techniques such as recurrent neural networks (RNNs) ~\cite{tkavcik2016neural,lin2017deep,song2016deeptransport, Feng:2018:DPH:3178876.3186058, liu2016predicting} to capture these dependencies in mobility patterns. A variety of data sources have been used  to drive the empirical studies as well as the subsequent modeling efforts, ranging from GPS, cellular, WiFi logs of mobile devices, social media check-in data ~\cite{becker2013human, Ashbrook:2003:UGL:945305.945310, tang2015uncovering, zheng2008understanding,hang2018exploring, Cho:2011:FMU:2020408.2020579}, as well as   transportation data such as taxi logs ~\cite{Ganti:2013:IHM:2493432.2493466,tang2015uncovering}.

{\bf \indent Outdoor versus Indoor Mobility:}
Much of the above work has focused on \emph{outdoor mobility} to understand how humans move from one location to another at the spatial scale of a city or community (i.e., at urban-scales) \cite{isaacman2012human,noulas2012tale, yuan2012discovering, song2016deeptransport,zhang2014exploring}.  Humans spend over 80\% of their lifetime indoors \cite{klepeis2001national}, and indoor mobility is known to be different from outdoor mobility patterns \cite{Zheng:2018:BAM:3276774.3276780,zhou2017mining}.
Specifically, indoor mobility is  concerned with how users and their devices exhibit nomadic behavior within buildings---that is, what locations (buildings, rooms) users visit, how long they stay at each location, and the transition path between locations; since indoor movements are based on walking, we are not concerned with the velocity of transitions---unlike outdoor mobility in vehicles, for instance.  


{\noindent \bf Mobility at spatial scales:} In the context of campus-scale mobility, mobility can be analyzed at two spatial scales:  inter- and intra-building scales.  Inter-building mobility is concerned with macro-scale mobility from one building to another; in this case, the \emph{entire} building is assumed to be \emph{single} spatial location and we characterize nomadic behavior in terms of time spent in a building, transition time to the next building, and so on.  Intra-building mobility is concerned with micro-scale mobility {\em inside}
a building--i.e., how the time spent in a single building can be further broken down into mobility across rooms or locations within that building; in this case, rooms, specific 
areas within the building or even access points are assumed to specific locations and mobility is viewed at the finer spatial scale across these locations.  
 Figure \ref{fig:hierarchical_view} 
illustrates the inter-building and intra-building trajectories of an actual user; at inter-building scale, the trajectory reveals the sequence of buildings visited over a day and the times spent in each, while at intra-building scale, the trajectory reveals what locations were visited by the user when visiting each of those buildings. 


{\bf \indent Multi-device Users:} Modern Internet users
carry multiple devices. 
A common use case is to own a smartphone and a laptop, but many users will own more than two devices that  include tablets, e-readers, and wearable smartwatches, among others.
Users will use their devices differently, causing them to exhibit different mobility patterns. For example, a user may carry their phone everywhere they go while at work, but they may not take their laptop to activities such as lunch that do not require the laptop. This will cause the mobile behavior of the laptop to deviate from that of the phone even though both are owned by the same user. Thus, we distinguish between {\em user mobility} and {\em device mobility} and assume that each mobile device exhibits a distinct mobility pattern, which approximates to varying degrees of the true mobility of their owner. 
Further, the user's device that exhibits the greatest mobility---often the one that the user takes with them "everywhere"---yields the best approximation of the user's mobility. Figure \ref{fig:traj_timeline} illustrates these differences by depicting mobile and laptop trajectory for an actual user. As shown, both devices visit the same location whenever the user brings both devices to that location; the figure shows that
the user often leaves the laptop at a location but takes the phone with them when visiting
other locations inside a building or even other buildings, causing trajectory deviations. The trajectories converge again when the user returns to the previous location.

{\bf \noindent Mobility in Campus Environments:} 
Campus settings are well-suited for studying  mobility for several reasons.  Users in university or corporate campuses spend a significant portion of the work day working or studying in such settings. Such users tend to be tech-savvy and own a multitude of mobile devices, and campus environments tend to have ubiquitous WiFi network coverage. Finally, since a campus comprises multiple buildings, it enables mobility analysis at different spatial scales, such as  intra- and inter-building characterization of mobility.

{\bf \noindent Relation to prior work on campus-scale mobility:}
While much prior work focuses on outdoor mobility, there have been a few 
campus-scale mobility characterization studies, 
at university campuses such as Dartmouth \cite{kotz2005analysis,Henderson:2004:CUM:1023720.1023739}, SMU \cite{Jayarajah15,Jayarajah18} and Tsinghua \cite{zhou2017mining} and in corporate campuses   \cite{balazinska2003characterizing}. 
   However, none of these studies have examined mobility of multi-device users or that at different spatial scales. These studies do not analyze
    mobility at micro- and macro-scales (nor are we aware of efforts to characterize outdoor urban-scale mobility at multiple spatial scales). Further, prior studies have not focused on users with multiple mobile devices---prior work from the  early 2000s was conducted in the pre-smartphone era and implicitly assumed  a single device per user environment, with laptops being the primary user device\cite{kotz2005analysis, balazinska2003characterizing}.  More recent studies \cite{Jayarajah15,Jayarajah18,zhou2017mining} did not focus on this specific research question, and have instead focused on other issues such as crowd activities \cite{zhou2017mining},  group behavior \cite{Jayarajah15,Jayarajah18} or networking aspects \cite{balazinska2003characterizing}. Table \ref{tab:prior} summarizes these differences.

%% file: dataset.tex
\section{Dataset and Methodology }
\label{sec:dataset}

This section describes our dataset, followed by the methodology used in our analysis.


\subsection{Campus WiFi Dataset}
\label{sec:wifi-dataset}
Our  campus-scale indoor  mobility study is based on 
 WiFi data from a large university campus (name removed for double-blind review). The campus comprises 156 buildings spread
over 1460 acres and is a residential campus where most undergraduate students live on campus. 
Wireless connectivity is available in all campus buildings and also in many outdoor spaces. The campus WiFi network consists of  5104 HP Aruba access points (APs) that are managed by seven wireless controllers. The controllers  receive syslog messages of all events seen by the APs; these logs contain many types of events, of which six events types are relevant to our study: (i) association, (ii) disassociation, (iii) re-association, (iv) user authentication, (v) deauthentication,  and (vi) drift events. Since the campus   WiFi network uses enterprise RADIUS authentication, all user devices must first authenticate themselves before they connect
to the network. Doing so generates authentication and deauthentication log messages, which allows the network to associate each device with a particular user. 
 Once authenticated, the device can then associate with a nearby access point, which
generates an association message in the event logs. If the device moves out of range or wakes up from sleep, it may generate disassociation, reassociation, or drift messages.


\begin{table}[ht]
\begin{minipage}[c]{0.5\linewidth}
     \centering
     \begin{tabular}{lc} \toprule
     Item Description & Value \\ \midrule
     Duration  & Fall 2018 (Sep-Dec)\\
     Num. events in log & 9.6 billion\\   
     Num. of Buildings & 156 \\
     Num. of APs &  5104 \\
     Num. of devices & 70,040 \\
     Num. of Student users & 24,791 \\
     Num. of Faculty-staff users & 5293 \\
     \bottomrule
     \end{tabular}
     \caption{Dataset Description}
     \label{tab:dataset_stats}
  \end{minipage}
\hfill
\begin{minipage}[c]{0.45\linewidth}
\centering
\includegraphics[width=2.1in]{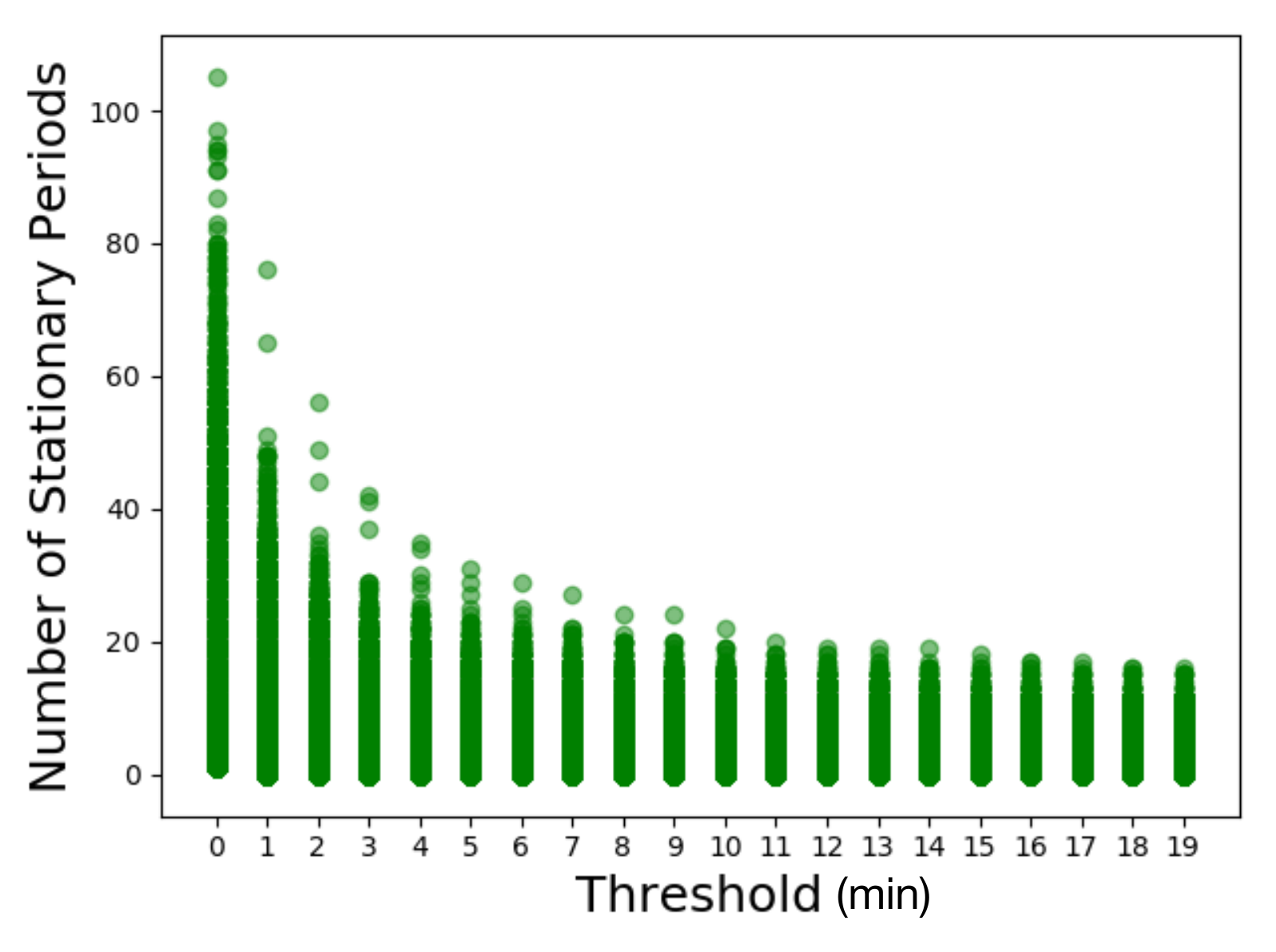}
\captionof{figure}{Impact of different thresholds for determining stationary periods.}
\label{fig:why_10mins}
\end{minipage}%
\end{table}


\noindent {\bf WiFi Logs:} Each event in the log consists of a timestamp, the event type, the MAC address of the device, and the Access Point ID. In addition, authentication and deauthentication events also include the user name and user type (which can be one of student, faculty-staff, or guest). For privacy reasons, all device MAC addresses and user names are anony\-mized using a SHA-1 hash function.  Since the location of all access points are known (in terms of the building and floor where they are deployed), each of six event types represents a {\em ``presence''} event, since it indicates the presence of that device in the vicinity of the access point (and its corresponding location).  The sequence of presence events generated by a device over the course of the day then reveals all the AP (and building-specific) locations visited by that device and the time spent at each location. 
Further, since each device must first authenticate to the network using its owner's user ID, the owner of each device is known, which in turn reveals the collection of devices owned by each user.
 
This data collection has been ongoing since 2013 for various longitudinal studies.  Unless specified otherwise, for computational tractability, our analysis focuses on a single university semester, namely Fall 2018, which spans from September to December 2018 (see Table \ref{tab:dataset_stats}). The event log for this one semester is over 260GB in size and contains 9.6 billion events spanning 70,040 devices, 24,791 student users and 5293 faculty-staff users.

 \noindent \textbf{Ethical considerations:}
 All data collection and analysis was conducted under a set of safeguards and restrictions approved by our Institutional Review Board  (IRB) and a Data Usage Agreement (DUA) with the campus network IT group.  All MAC addresses and user names in the trace are anonymized using a strong hashing algorithm. Further, the software used to perform  anonymization  
performs all  hash-based anonymization prior to storing the data on disk (i.e., stored files only contain anonymized  data) and the hash keys are known only to an IT manager and unknown to us.
Our IRB protocol and DUA explicitly prohibit us from any mobility analysis that could de-anonymize users.
 Researchers need to undergo ethics training and 
sign a form consenting to these restrictions on the data. 
Finally, all end-users who connect to the campus WiFi network need to consent to the campus IT agreement that allows syslog data to be collected for security reasons and for aggregated
analysis.  

 
\subsection{Trajectory Extraction Over Noisy Data} 
We now describe our methodology for extracting the mobility trajectory exhibited by each device using noisy syslog traces and validation of the generated trajectories. 
From the overall trace, we first extract a trace of events generated by that device using its hashed MAC address.  This device-specific trace enables us to reconstruct the trajectory of the device over each day. Specifically, the timestamped presence messages reveal the sequence of locations visited by the device during each day and the time spent at each location.  For students who stay on campus in residence halls, we can reconstruct a trajectory for the entire day, while for faculty, staff and student who live off-campus, we can reconstruct device trajectories for the portion of each day spent on campus. Each device trace reveals a trajectory comprising a sequence of stationary and transition periods. A \emph{stationary period} indicates that the device is stationary at a particular location (by virtue of being associated with the same AP for a period of time). A \emph{transition period} indicates that the device is "on the move," which is seen as a sequence of association events with different APs, each of a short time duration.  

For the purpose of this study, we use a 10 minute threshold to distinguish between a stationary and transition state for a device---if a device is associated with an AP for a duration greater than 10 minutes, we label the current location as a stationary period, otherwise the location is assumed to be part of a transition period in the overall device trajectory.  The 10 minute threshold was chosen after a careful analysis of the data. Figure \ref{fig:why_10mins} depicts the number of stationary periods (i.e., locations) visited by a device obtained for different thresholds. A smaller threshold implies that even short stays at an access point will get labeled as stationary periods. The figure depicts that the curve flattens at 10 minutes and stays flat beyond this threshold value; such a 10 minute threshold, also employed by others~\cite{kim2006extracting},  aligns with human notions of visiting a location versus transiting through one. 

\noindent {\bf Handling Noise:} The trajectories extracted from raw traces will be inherently noisy.
For example, mobile devices may connect to access points that not the most proximal, or 
``ping-pong'' between nearby access points even though the user is stationary. Similarly,
when the user is walking to a new location, devices may connect to distant APs in weakly connected regions or exhibit similar ping-pong switching effects. Since we are using the AP location to determine the device location, all of these effects introduce noise or spurious location changes into our extracted trajectories. To address this issues, the raw
traces are subjected to a multiple
 filtering and smoothing steps during trajectory extraction to remove noise and obtain clean trajectories for each device. Full details of these smoothing and filtering steps for data cleaning may be found in Appendix \ref{sec:smoothing}.    

\noindent {\bf Validating Trajectories:} We conducted a small-scale  study 
to validate that device trajectories derived from the WiFi dataset corresponds to the ground-truth device trajectory. To do so,
	we had volunteers mimic user behavior by walking with a phone to various campus buildings, spend some time inside each building, and then walk to next building and so on. The ground truth trajectory (i.e., locations and times) recorded by the user were compared to that extracted from the WiFi log of the phone. 
Our  validation study revealed more than 99\% accuracy between the extracted locations and the  ground truth for indoor locations and deviations of no more than 20-40 meters in outdoor locations when walking outside, providing confidence that our dataset enables us to study campus-scale mobility  behavior.  Additional 
 details of the validation study may be found in Appendix \ref{sec:validate}.

%% file: dev_classification.tex
\subsection{Device Classification}
\label{sec:dev_classification}

Our WiFi logs allow us to associate a device to a user and determine all devices belonging to a user, but they do not include any information to determine the \emph{type} of each device---anonymized MAC addresses alone do not reveal device type.\footnote{For privacy reasons, even partial MAC address prefixes, which reveal vendor and device type, are unavailable to us.}  Consequently, we develop a simple classification technique that uses the network behavior exhibited by each device to infer its device type. 


First, we observe that differences in OS power management results in different network behavior during idle periods for different device types. Devices such as phones, and many tablets, tend to be powered on at all times---even when the user
is not actively using the device. When these devices become idle, their network interfaces enter a low-power listen state but stay powered up (e.g., to receive push notifications, chat messages, or video calls). Hence the device continues to maintain a network presence and is periodically visible to the WiFi network.  Consequently, when a user walks from one location to another, the access points along the path periodically see the presence of the device (through scans, association or disassociation messages). Of course, if the user actively uses their device when walking, the device maintains a continuous, rather than periodic, network presence along the path. Either way, the trajectory of such always-on devices is visible to the network during a location transition.

In contrast, mobile devices such as laptops tend to hibernate when not in active use (e.g., when the laptop lid is closed). The hibernate power state results in network interfaces being powered down, and the device no longer maintains a network
presence while hibernating. Consequently, when a user walks with a laptop
from one location to another, the device is not visible to the network during the walk
and only becomes visible when a user begins using the device at the new location. 

This difference in network behavior during transition periods and the resulting network visibility of the device (or lack thereof) enables us to distinguish between, and classify, \emph{always-on} and \emph{hibernating} devices. The most common always-on device in our current environment is a phone\footnote{Smart watches, which are another type of always-on device, were not present in our dataset since they do not support   RADIUS authentication for enterprise WiFi networks.} 
while the most common hibernating device is a laptop.

\noindent \textbf{Validation:} Since the above classification method is a heuristic, we conducted
a small-scale study to validate its accuracy. We had a volunteer mimic actual user behavior 
by visiting various buildings and walking within and across buildings over a period of 3 weeks and 8 hours each day; the user used two iOS devices (phone and tablet), one android phone, and 3 laptops with MacOS, Windows  and Linux. In each case, we used the daily trace from the device to classify its type using the 
above heuristic. We found that all devices get classified with perfect accuracy as an always-on or hibernating device when using a day long trace. The only way a laptop can get mis-classified as an  always-on device is to walk with the laptop's lid open so that it shows network presence during location transitions.  However, it is highly unlikely that users will {\em always} use a laptop in this manner over multiple days, even if they occasionally walk to a location with an open laptop.
Further, our analysis uses day-long traces over multiple randomly-chosen days from the 4 month dataset to classify each device, which significantly reduces the changes of mis-classification. Consequently, our heuristic is able to classify devices as always-on or hibernating with high accuracy.



\begin{table}
    \begin{minipage}{.45\textwidth}
      \centering
      \begin{tabular}{cc}
        \toprule Device Characteristics & \% Devices\\
     \midrule Hibernating  & 26.91\% \\ 
              Always-on  & 73.08\% \\
        \bottomrule
      \end{tabular}
    \caption{Distribution of always-on and hibernating devices.}
    \label{tab:tab1}
    \end{minipage}
    \hspace{0.6cm}%
    \begin{minipage}{.45\textwidth}
      \centering
      \begin{tabular}{cc}
        \toprule  Device Type & \% Users \\
    \midrule Primary  & 89.54\% \\
          Secondary  & 94.52\% \\
        \bottomrule
      \end{tabular}
      \caption{Ownership of primary and secondary devices.}
    \label{Tab:tab2}
    \end{minipage}
\end{table}

\begin{figure*}[t]
\begin{minipage}[c]{0.3\linewidth}
\includegraphics[width=\textwidth]{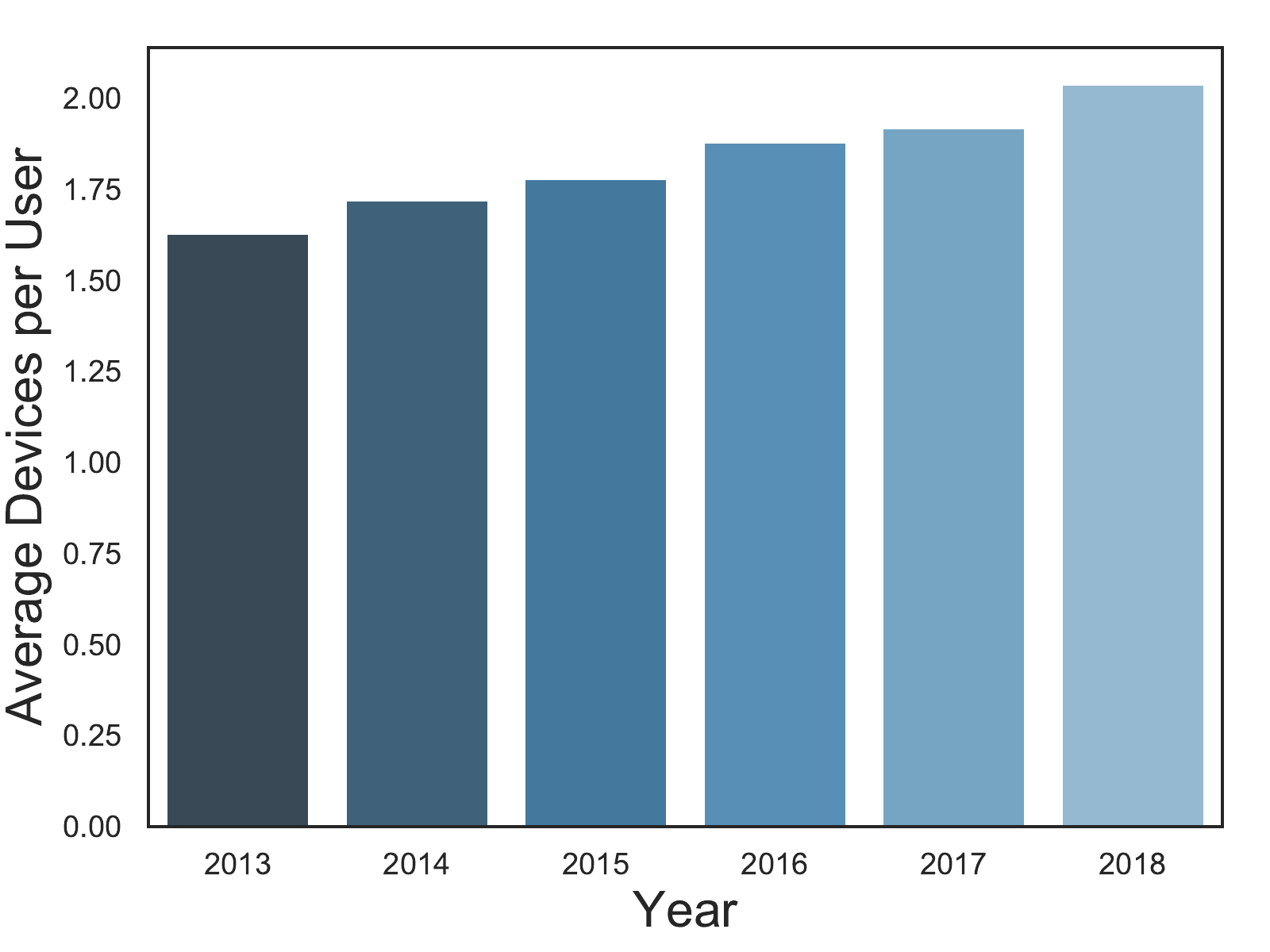}
\caption{Device ownership has grown steadily since 2013.}
\vspace{-0.05in}
\label{fig:trend}
\end{minipage}
\hfill
\begin{minipage}[c]{0.3\linewidth}
    \includegraphics[width=\textwidth]{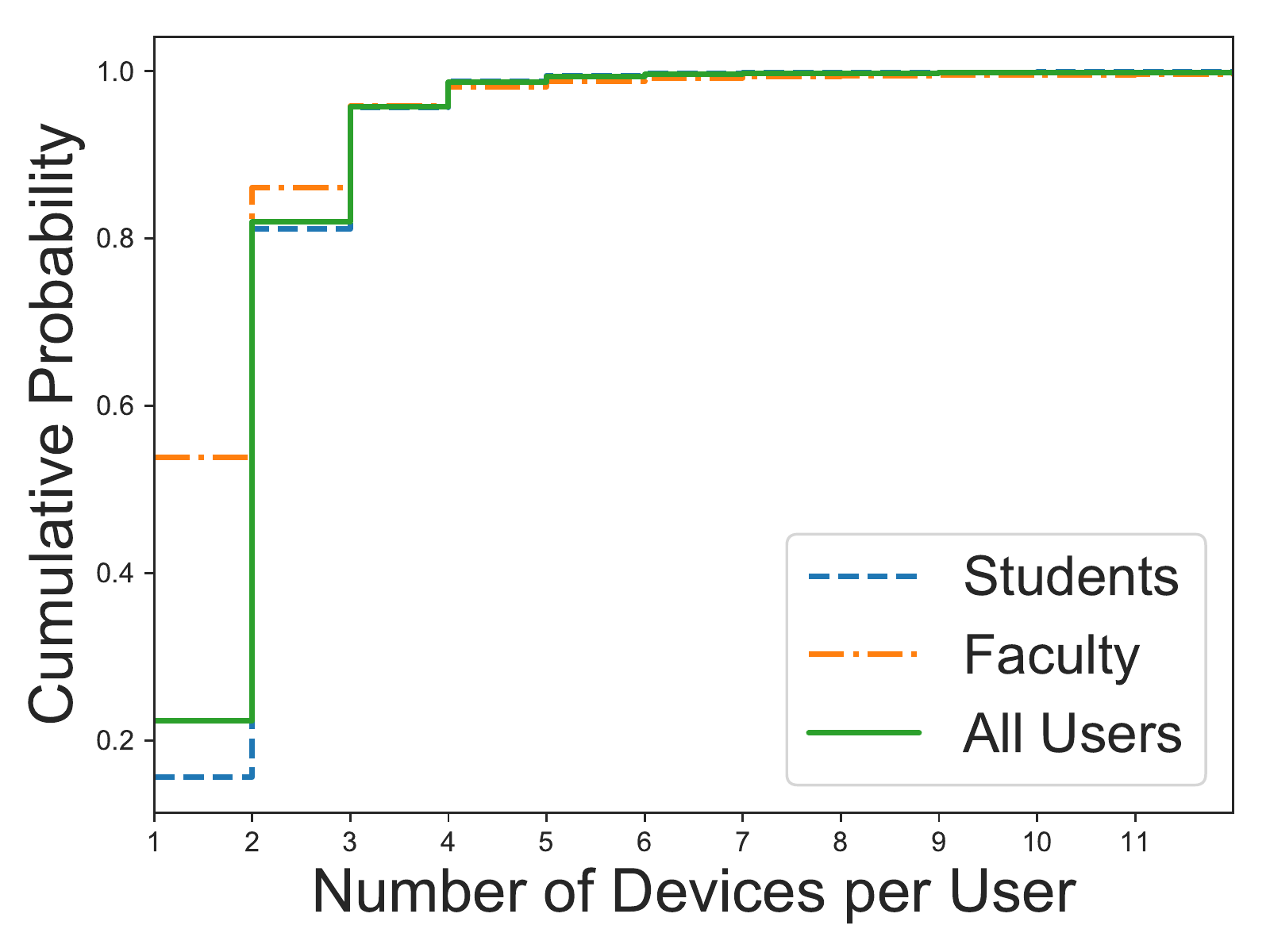}
    \caption{CDF of the number of devices owned per user.}
\vspace{-0.05in}
\label{fig:device_ownership}
\end{minipage}
\hfill
\begin{minipage}[c]{0.3\linewidth}
\includegraphics[width=\textwidth]{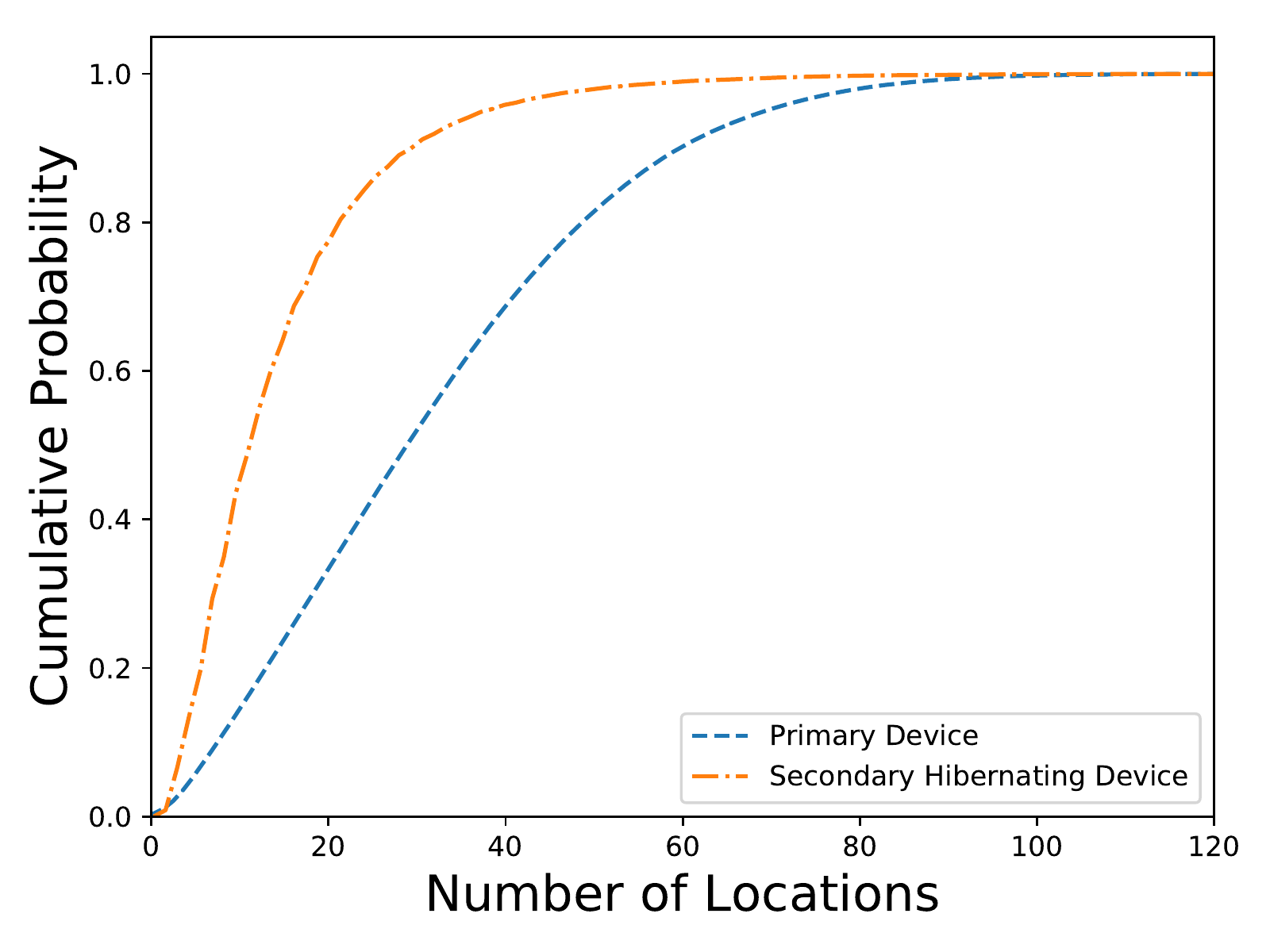}
\caption{CDF of number of locations visited by device type.}
\vspace{-0.05in}
\label{fig:AP_stationary}
\end{minipage}%
\end{figure*}


\noindent \textbf{Result:} We classified all devices in our WiFi log using the above method, and Table \ref{tab:tab1} depicts our results.
  As shown, 73.08\% of all devices maintain a network presence during location transitions and are classified as always-on devices. The remaining 26.91\% 
 devices do not exhibit any presence during location transitions and are classified as
 hibernating devices.

For the purpose of our study, we further classify all devices belonging to each user as
primary and secondary.  A  user's {\em primary} device is defined as the always-on device that exhibits the greatest mobility (greatest number of stationary periods per day) across all always-on devices owned by that user.  All other devices belonging to that user,   whether always-on or hibernating, are defined as \emph{secondary} devices.  By virtue of being the user's most mobile device, the primary device also provides the best approximation of the user's actual mobile behavior.
With a high likelihood, a user's primary device is likely to be a mobile phone.  Further, with high likelihood, a hibernating secondary device belonging to the user is likely to a laptop.  After labeling all devices as always-on and hibernating and then labeling each user's device as primary and secondary, we see in Table \ref{Tab:tab2}  that 89.54\% of users own a primary device. This also implies that 10.45\% of our users either do not own a smartphone or do not connect their phone to the campus WiFi network.
The table shows that  94.52\% of our users own at least one secondary device.  
Since multi-device ownership is common on our campus, there is a substantial overlap between these two user groups, as discussed next. 

%% file: multi_device_users.tex
\section{ Multi-device Users}
\label{multi_device_users}

In this section, we analyze the mobility of different devices owned by a user and the  mobility across device types.

\input{devices_users}

\subsection{Characterizing Device Mobility}
\label{sec:device_mobility}

To characterize the mobility of different types of devices owned by a user, we considered
the primary device (i.e., the phone) and the hibernating secondary device (i.e., the laptop) for each user, which is also the common case for device ownership on our campus.

Figure \ref{fig:AP_stationary} depicts the CDF of intra-building locations visited per day for the two device types. The CDF reveals that phones visit 35.9 intra-building locations, on average, per day, while laptops visit 10.2 locations, which yields a 3.5X more mobility for phones than laptops. More generally, since  devices as laptops are less portable than phones, users will carry them to fewer locations, causing them to exhibit lower mobility---the ``less portable'' the device, the lower its mobility.  In the future, as wearable devices such as smartwatches become common, we expect them to be even more mobile than today's phones. 

\begin{figure}[h]
\includegraphics[width=2.5in]{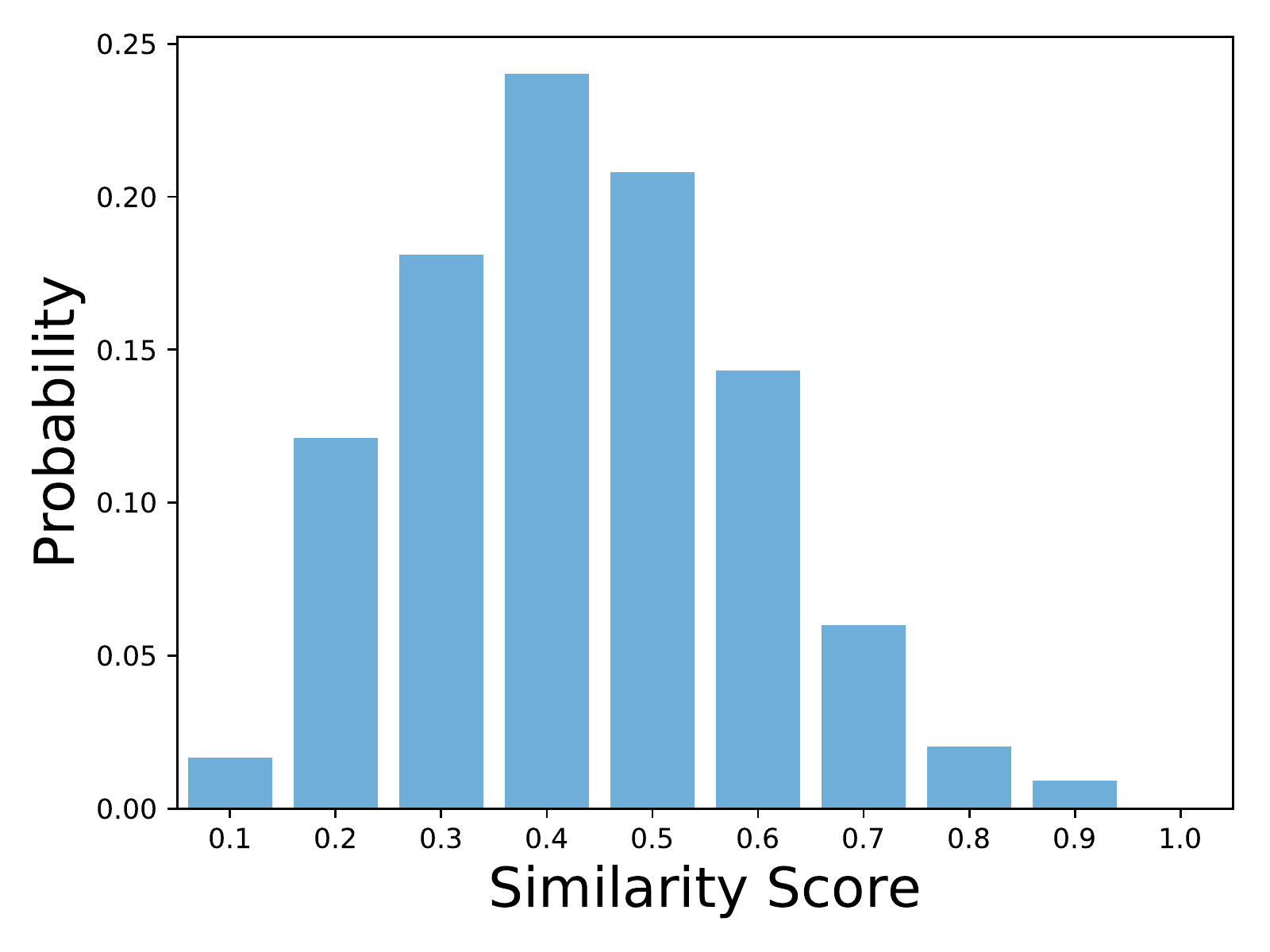}\\
\vspace{-0.2in}
\caption{PDF of similarity scores of  primary and secondary device trajectories.}
\vspace{-0.05in}
\label{fig:ps_sim_score}
\end{figure}

Recall also from Figure \ref{fig:traj_timeline} that while the phone trajectory will often deviate from the laptop and the two will converge again when the user returns to the laptop's location. Thus, despite having lower mobility, the laptop's trajectory is correlated to the phone's trajectory since both depend on the user's mobility behavior. 
To understand the degree of similarity between the two, 
we computed the pairwise similarities in the stationary location trajectories for each user's phone and laptop using Longest Common Subsequence (LCSS) score.
We choose LCSS as a measure of similarity since it is robust to noise and can handle synchronous or random shifts of the location sequence \cite{wang2013effectiveness}. Thus, small variations in trajectories do not have a large impact on the similarity measure.
We compute the similarity score as a ratio of the length of LCSS to the length of the union of the primary and secondary trajectories; the higher the score the higher is the device trajectory co-relation and vice-versa. Figure \ref{fig:ps_sim_score} depicts the PDF of the similarity scores obtained for all users. The similarity scores range from 0.06-0.86, with a mean of 0.37. The plot shows that 31.9\% device pairs have a weak similarity score of less than 0.3, 59.1\% device pairs have a moderate similarity score between 0.3 and 0.6 and 9\% device pairs have a high similarity score of 0.6 or more.  Thus, more than two-thirds of the users use their phones and laptops such that the two device trajectories show moderate to strong correlations, and this behavior is true despite phones having 3.5X higher mobility than laptops. The maximum similarity score is 0.86 which indicates that {\em even the most correlated pair of devices nevertheless see some dissimilarities in their trajectories.}  Our analysis shows that a user's mobile devices should not be viewed as independent due to their moderate to strongly correlated mobility patterns. Further, these correlations vary significantly across users, which should also be considered in system design and modeling.

\begin{mdframed}{
        \textit{ \textbf{Key Takeaway}}
        \begin{itemize}
            \item More portable devices such as phones exhibit 3.5X more mobility in terms of location visits than laptops. 
            \item  Primary and secondary device trajectories for over two-thirds of the users show moderate to strong correlations.  
        \end{itemize} 
    }
\end{mdframed}

\input{similarity}

%% file: devices_users.tex
\subsection{Multi-device Ownership}
\label{sec:devices_users}

To analyze device ownership, we first consider our entire longitudinal dataset spanning 2013 to 2018. Figure \ref{fig:trend}
shows the mean number of devices per user over this five year period. As shown, device ownership has grown steadily in recent years---the mean number of mobile devices per user grew from 1.63 in 2013 to 2.04 in 2018. Next, we focus on the Fall 2018 semester used for this study and analyze multi-device ownership across broad classes of users. 
Figure ~\ref{fig:device_ownership} plots the probability distribution of device ownership across students, faculty-staff, and the combination of the two.  
The CDF shows that the majority of the users---84.4\% of the students and 46.1\% of faculty-staff---own two or more devices. 
The average student owns 2.1 devices, while the average faculty-staff user owns 1.7 devices, indicating that students own more devices, on average than other user types. This is not surprising since younger users tend to be more tech-savvy, and furthermore, a large majority of students ($> 60\%$) stay on-campus and connect all of their devices to the networks (while faculty and staff who live off-campus may not bring all of their devices to work). Interestingly, the figure also shows that  18\% of users own three or more devices, with  1.33\%  users owning five or more devices.

\vspace{0.05in}
\begin{mdframed}{
        \textit{ \textbf{Key Takeaway}}
        \begin{itemize}
            \item Device ownership has increased steadily over time, and the typical user now owns 2.04 devices. 
            \item 18\% of the users own three devices or more.
        \end{itemize}  
    }
\end{mdframed}

%% file: dev_hierarchical_mobility.tex
\section{Macro and Micro-scale Mobility}
\label{dev_hierarchical_mobility}

In this section, we analyze  mobility at macro (inter-building) and micro (intra-building) spatial 
scales. Unless specified otherwise, all results in this section are based on the users' primary
device. 

\input{inter_bldg_analysis}

\input{intra_bldg_analysis}

%% file: inter_bldg_analysis.tex
\subsection{Macro-scale Inter-building Mobility}
\label{sec:interbldg_analysis}

To analyze mobility at the macro scale of entire buildings,  consider the trajectory of each user's primary device, which is a  sequence of AP locations visited by that user. Since building- and floor-specific locations
of each AP are known,  we can assign a building label to each visited AP, and then aggregate a consecutive sequence of APs with the same building label as a \emph{single} location, representing a visit to that building with a corresponding aggregated visit duration. The transformed
inter-building trajectories then yield a sequence of buildings visited by each user, time
spent in each building, and transitions across buildings.  At this macro scale, user trajectories are only concerned with visits to buildings and transitions between buildings, but not what happens inside a building. 

\begin{figure}[t]
\begin{center}
    \begin{tabular}{cc}
    \includegraphics[width=0.45\linewidth]{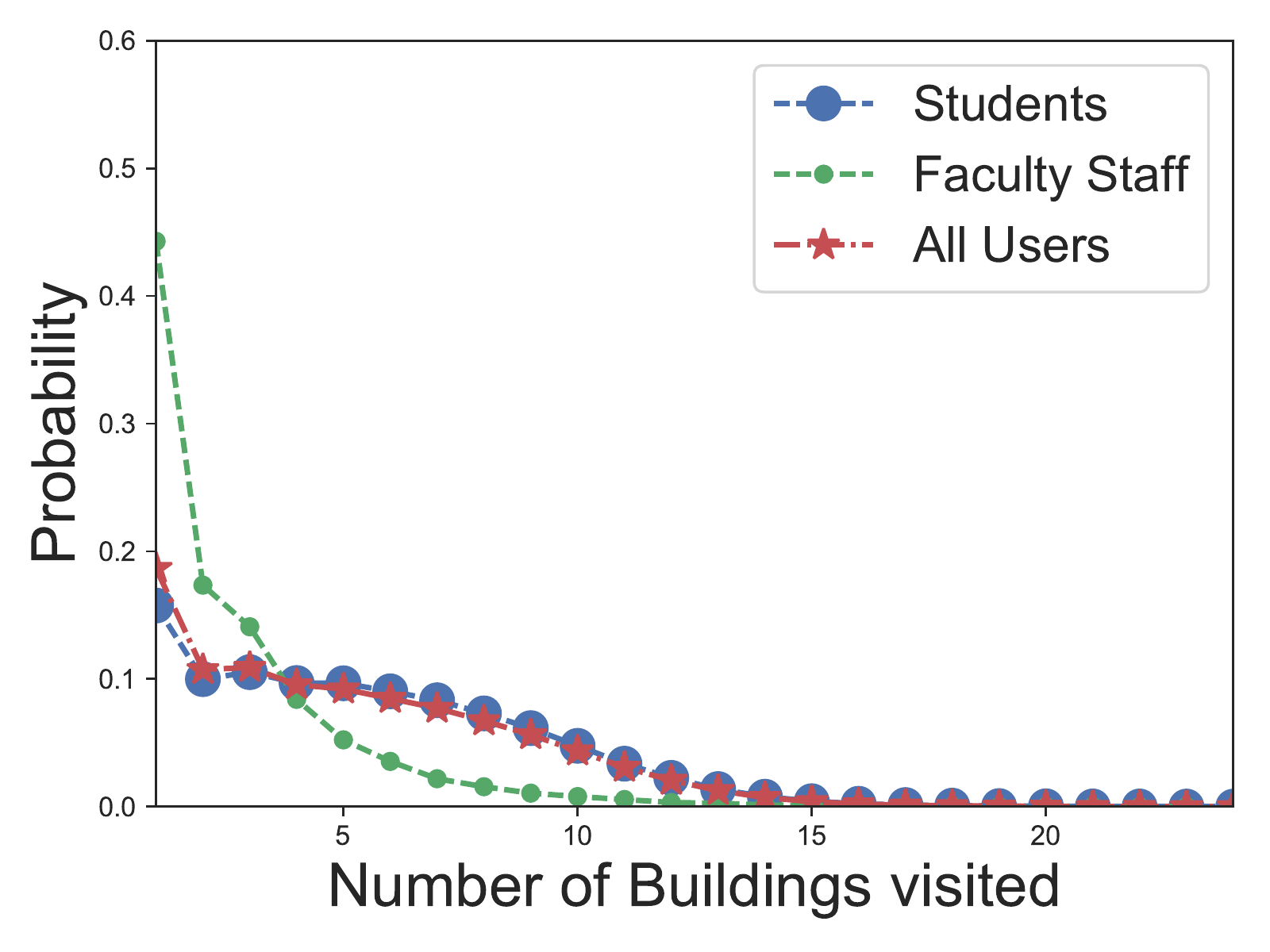} &
    \includegraphics[width=0.45\linewidth]{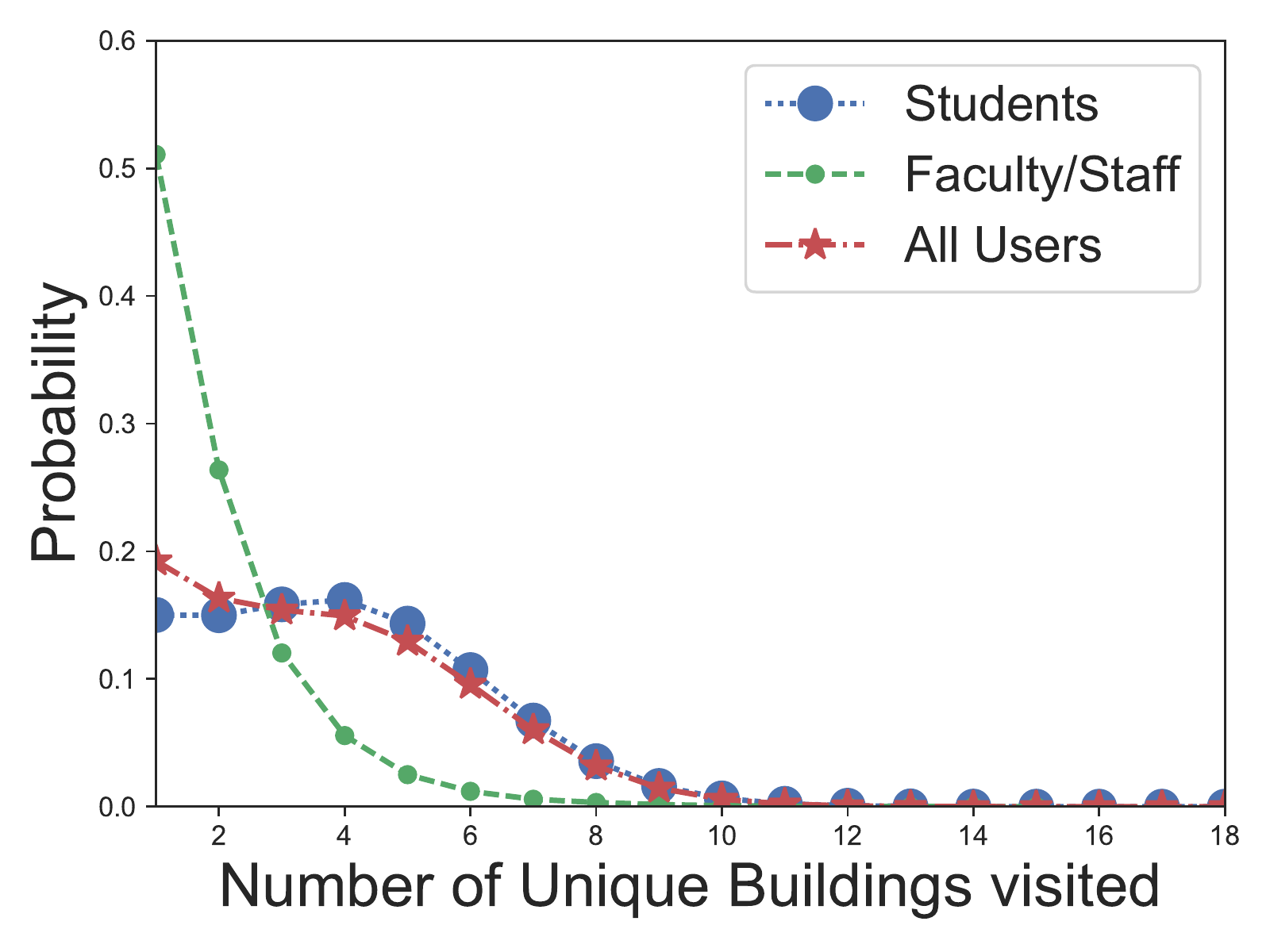}\\
    (a) Buildings visited & (b) Unique building visited 
\end{tabular}
\end{center}
    \caption{Distribution  of the number of buildings visited per day by campus users.}
\vspace{-0.05in}
\label{fig:LoI_visited}
\end{figure}

\begin{figure}[t]
\begin{center}
\begin{tabular}{cc}
    \includegraphics[width=0.45\linewidth]{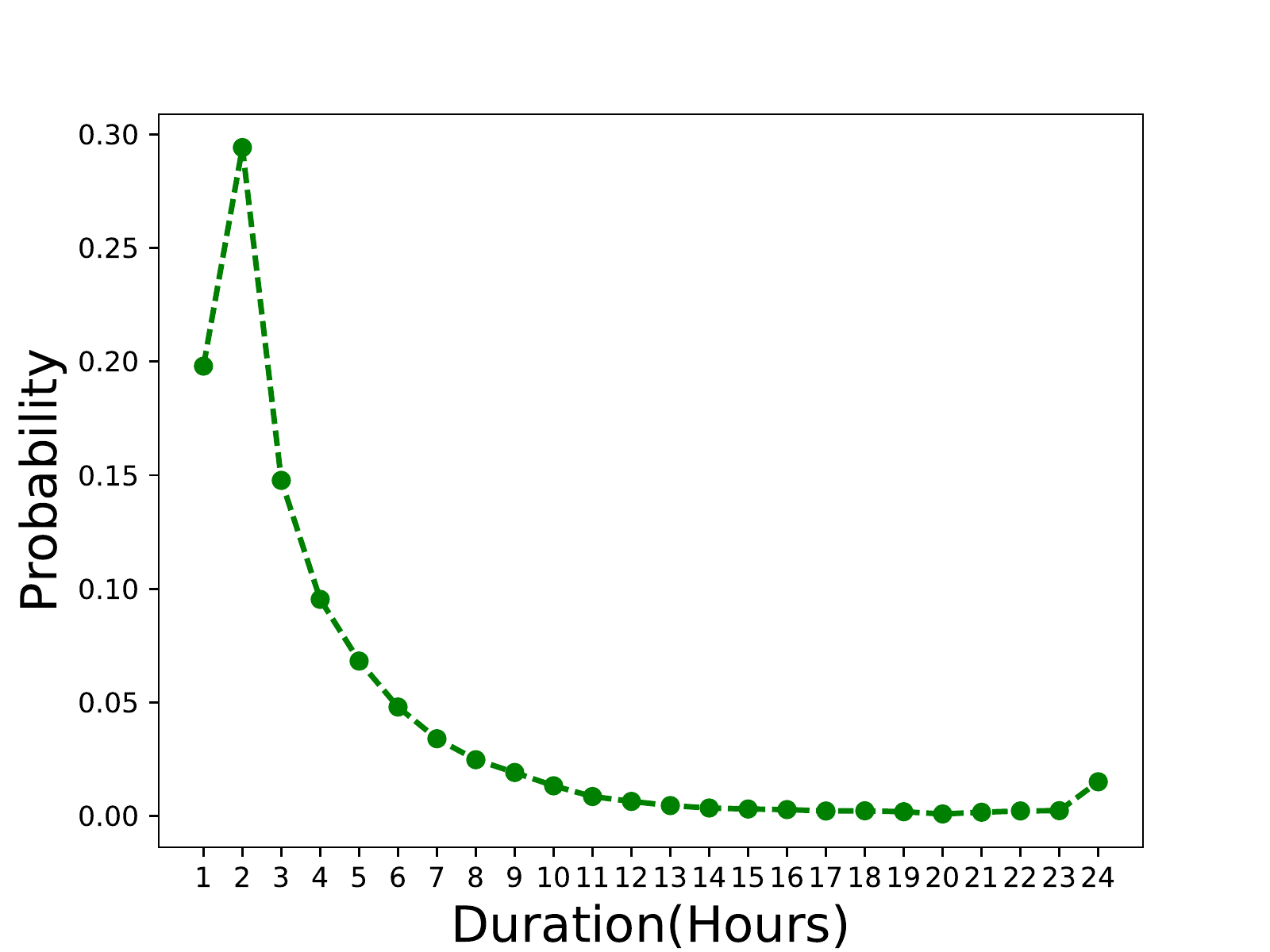} &   
    \includegraphics[width=0.45\linewidth]{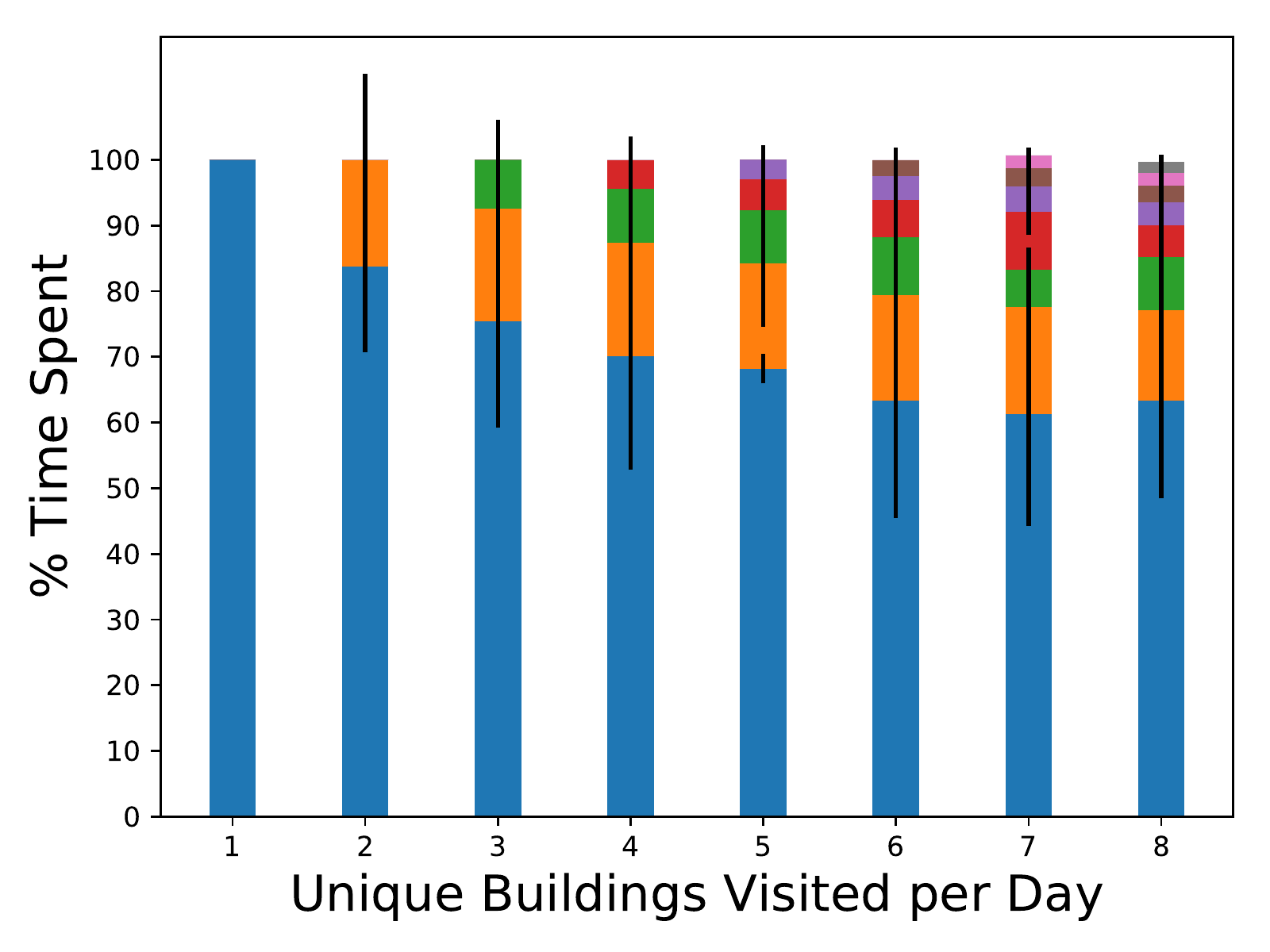}\\
	(a) PDF of time spent in a building & (b) Distribution of time spent across buildings
\end{tabular}
\end{center}
\caption{Distribution of time spent by users in campus buildings}
\label{fig:time_dist}
\end{figure}

{\em Stationary period analysis:}
Figure \ref{fig:LoI_visited}(a) plots the distribution of the number of buildings visited by a user  
over a day. The distribution reveals that the average user visits 4.1 buildings per day; highly
mobile users, depicted by the 90-th percentile of the distribution, visit 9.8
buildings per day. The distribution also shows that students are more mobile than faculty and staff, with students visiting 4.4 buildings per day, on average, and faculty and staff visiting 1.2 buildings,  on average. 
Figure \ref{fig:LoI_visited}(b)  plots the distribution of the {\em unique} number of buildings visited by users each day (where multiple visits to a building count as a single unique location). The figure shows that
a user visits 2.7 unique buildings, on average, each day, which implies users often return to their primary office building after a visit to another building or visit the same building (e.g., dining hall) multiple times per day. 

Figure \ref{fig:time_dist}(a) shows the PDF of the time spent in each building by a user. 
The PDF shows that a campus user spends 109 minutes, on average, visiting a campus building. Further analysis of this distribution reveals that about 30\% of building visits last less than 1.5 hours; 29\% of all building visits are long visits, lasting an average of 5.8 hours, indicating that the tail of the distribution has a substantial mass. 
Figure \ref{fig:time_dist}(b) plots the total time spent in each unique building visited versus the number of buildings visited by users each day.  The figure shows that both less mobile as well as highly mobile users spend between 60 to 80\% of their day in a single building, with the remainder of the day spent visiting other buildings for shorter periods. This result shows that most users spend a majority of their  day in a single ''home'' building (e.g., office or residence hall)

\begin{figure}[h]
\begin{center}
\includegraphics[width=0.54\linewidth]{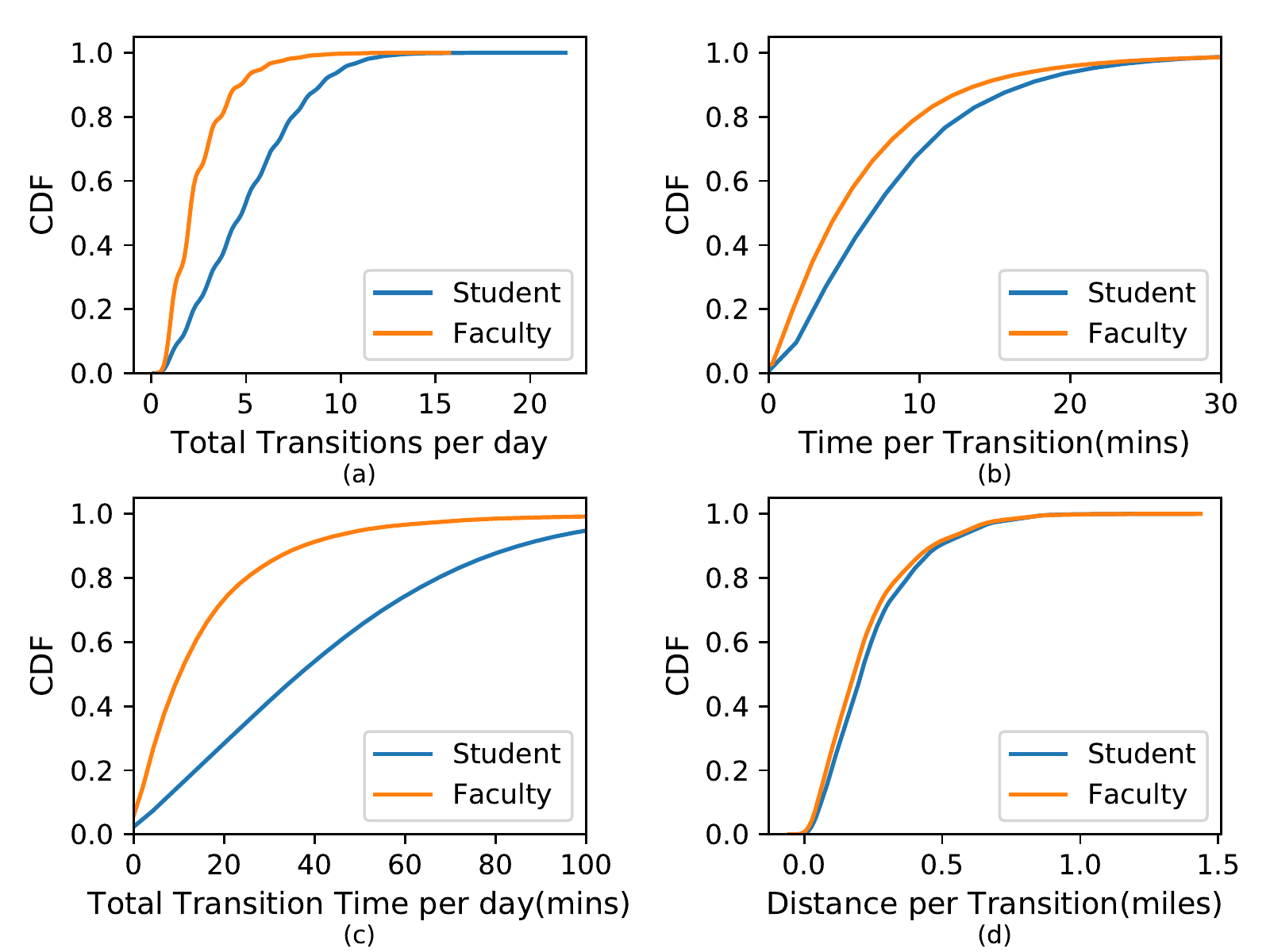}
    \caption{Inter-building mobility analysis: (a) CDF of transitions per day (b) CDF of duration of each transition (c) CDF of total transition time per day (d) CDF of mean distance traveled per day}
\label{fig:inter_bldg_transitions_stats}
\end{center}
\end{figure}

{\em Transition analysis:}
Next, we analyze temporal aspects of inter-building mobility by focusing on inter-building transitions.  Figure \ref{fig:inter_bldg_transitions_stats}
(a) depicts the CDF of the number of transitions per day made by campus users.  Since each visit to a building must be preceded by a transition, the mean number of transitions is the same as (or, for off-campus users, one more than) the number of buildings visited, with a mean of 4.1 transitions per user per day.
Figure \ref{fig:inter_bldg_transitions_stats}(b) depicts
the CDF of the duration of each inter-building transition. The figure shows that average transition time, which is usually a walk between two buildings,  lasts 8.4 minutes and 6.5 minutes for students and faculty-staff users, respectively. Figure  \ref{fig:inter_bldg_transitions_stats}(c) shows the CDF of the total time spent
in walking between buildings over the entire day. The figure shows that the average student spends 42.2 minutes per day walking between buildings, while the average faculty-staff user spends 16.1 minutes. Finally, Figure  \ref{fig:inter_bldg_transitions_stats}(d) depicts the CDF of the distance traveled when walking from buildings to another, and shows the average
walk between campus building is 0.22miles. 

Interestingly, we also find that about 15\% of all inter-building transitions on campus are loops, with the same origin and destination.  Such transitions last 15 minutes, on average, and involve a walk of 0.6 miles. We believe that such transitions occur when users go for a walk during break, or walk to another building to for an errand and return to their previous building. 

\begin{figure}[h]
\begin{center}
\begin{tabular}{cc}
    \includegraphics[width=0.45\linewidth]{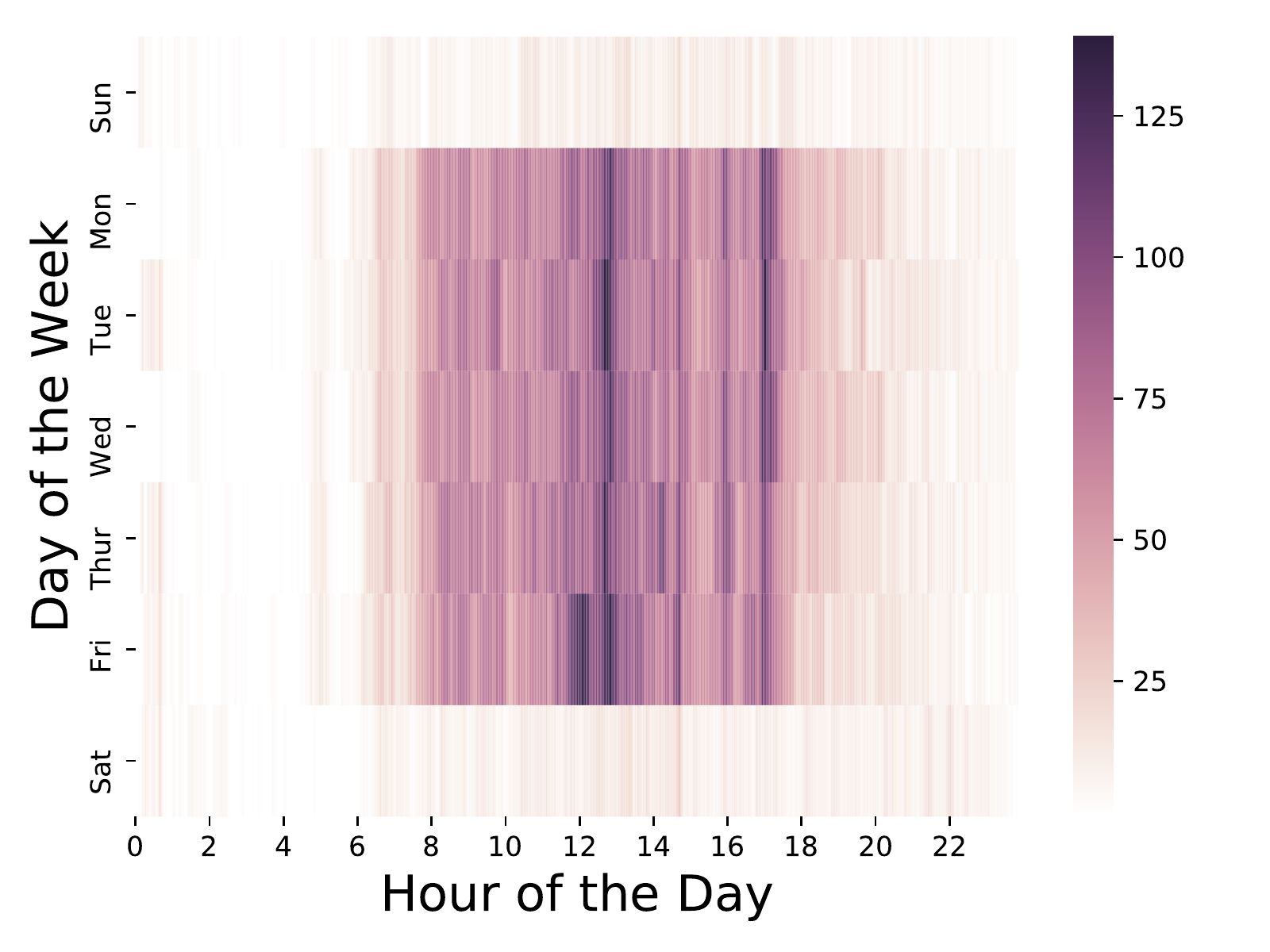}&
    \includegraphics[width=0.45\linewidth]{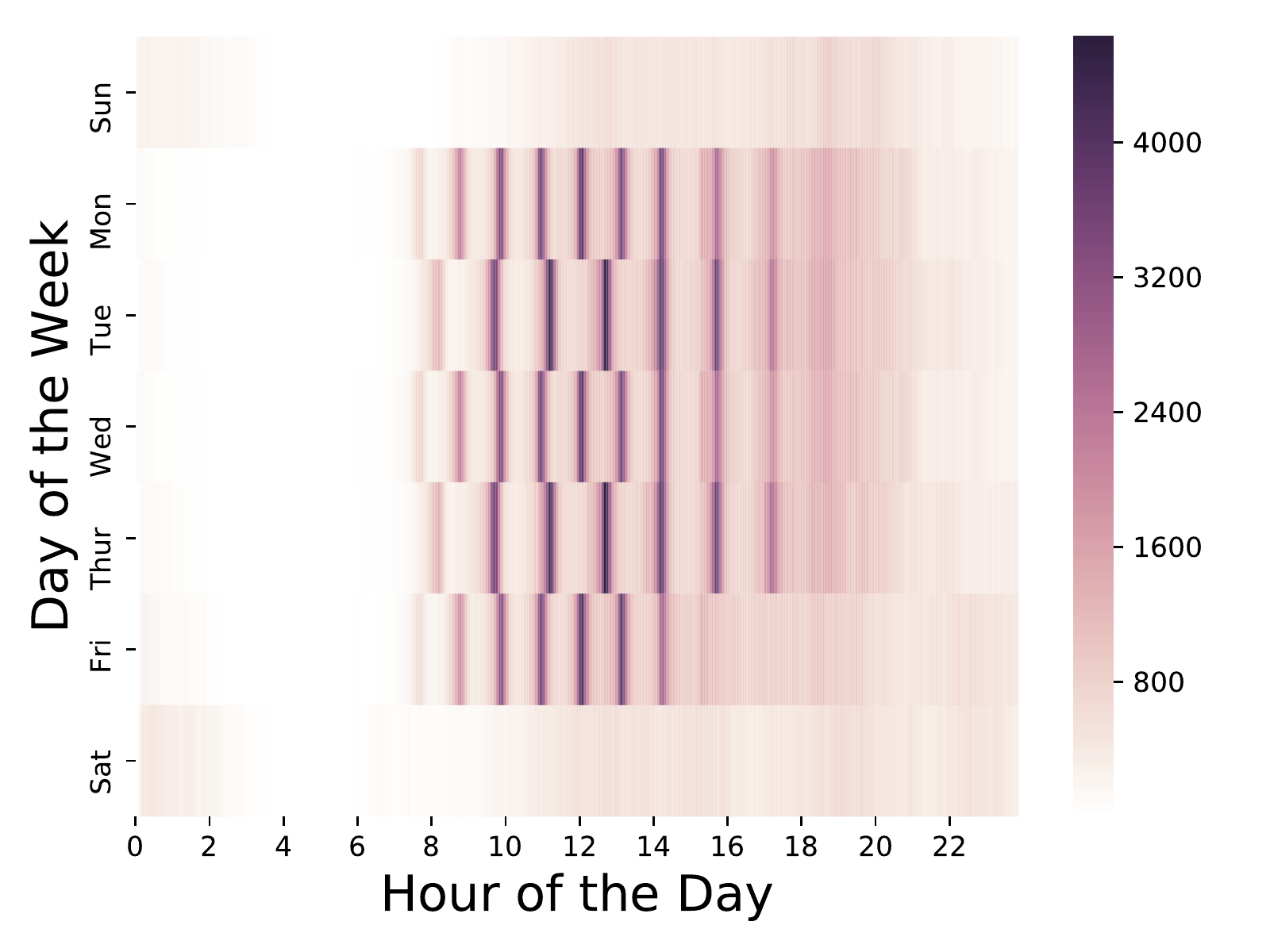}\\
(a) Faculty-staff  & (b) Students
\end{tabular}
\end{center}
\caption{Heatmap of inter-building transitions. Student mobility is aligned with class  and meal times, while faculty and staff mobility is more dispersed with a cluster around the lunch hour.}
\vspace{-0.05in}
\label{fig:inter_bldg_transitions_t}
\end{figure}


Figure \ref{fig:inter_bldg_transitions_t}(a) and (b) depicts the heatmap of the 
{\em when} users move between buildings. 
Faculty and staff users
make inter-building visits at all times of the day during working hours, as shown in Figure \ref{fig:inter_bldg_transitions_t}(a), with a
significant density of transitions during the noon lunch hour. 
Transition times during evenings, nights, and weekends are more diffused for these users.
In contrast,
Figure \ref{fig:inter_bldg_transitions_t}(b) shows that student transitions during weekdays are highly aligned with lecture start and end times between 8 to 6pm, and are more dispersed during other hours.
The weekend transitions do not show such
patterns since there are no classes on weekends. 

\vspace{0.05in}
\begin{mdframed}{
        \textit{ \textbf{Key Takeaway}}
        \begin{itemize}
          \item Users visit 4.1 buildings per day and spends nearly 2 hours, on average, in a building. 
          \item Highly-mobile users visit 2.4X more buildings than the typical user. 
          \item A third of the building visits are short, lasting 90 min or less, while a third of the visits last 5.8 hours, indicating the tail of the distribution has a significant mass.
          \item Even in a large campus, users tend to show a primary affinity to a single building,
           spending over 60\% of their day in that building. 
          \item A typical transition (walk) between buildings lasts 8.1 minutes;  the timings of such transitions is strongly correlated with class and meal times during daytime hours.  
        \end{itemize} 
        }
\end{mdframed}

%% file: intra_bldg_analysis.tex
\subsection{Micro-scale Intra-building Mobility}
\label{sec:intra_bldg_analysis}

Next, we examine intra-building mobility by characterizing micro-scale behavior of 
what users do while inside a building. We do so by analyzing trajectories of AP locations visited by the user's primary device inside each building.

\begin{figure}[h]
    \includegraphics[width=0.9\linewidth]{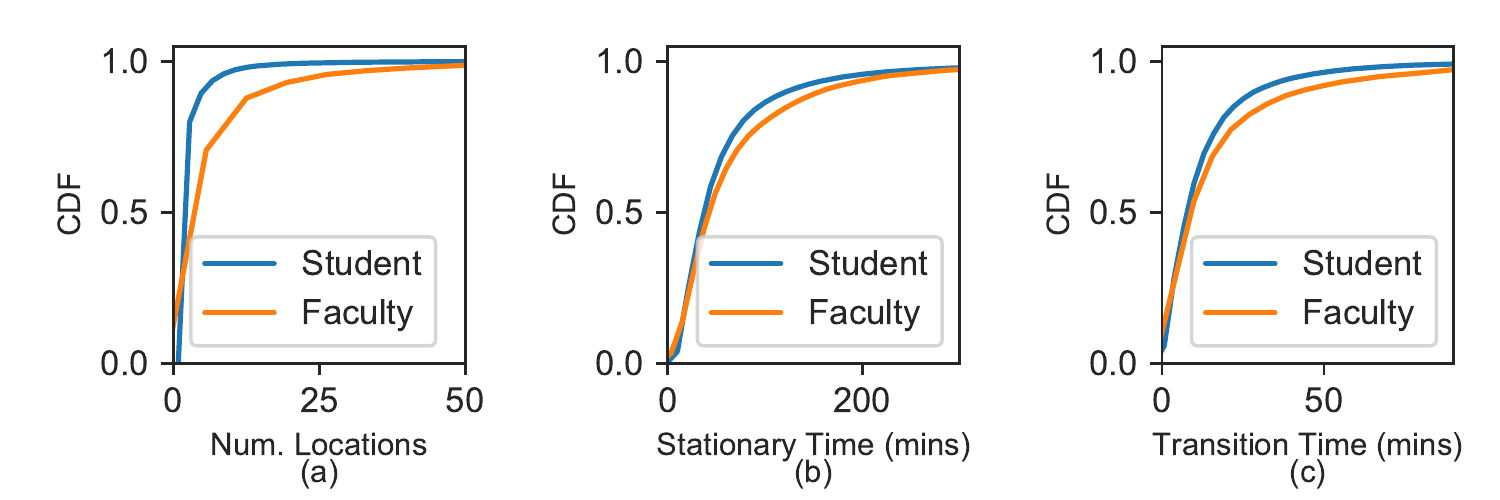}
    \caption{Intra-building mobility analysis: (a) CDF of number of locations visited, (b) CDF of time spent at each location, and (c) CDF of intra-building transition times. 
    }
\label{fig:pri_sec_stat}
\end{figure}



Figure \ref{fig:pri_sec_stat}(a) shows the CDFs of locations visited (i.e., stationary periods) by a user inside a building. Interestingly, our results show that students visit 8.6 locations, on average, inside a building, while faculty and staff users visit 12.1 locations.  In other words, at intra-building scale, faculty and staff exhibit higher mobility (by 1.4X) than students, which can be attributed to them spending more time
inside each building due to the {\em lower} inter-building mobility. 
 Since each stationary period is preceded by a transition,  Figure \ref{fig:pri_sec_stat}(a) also represents the mean number of transitions inside a building (not counting the final transition when the user departs from the building). 
Figure \ref{fig:pri_sec_stat}(b) shows the CDF of time spent at each location inside a building.  The figure shows that students and faculty spend 37 and 40 minutes, respectively, on average, when visiting a location inside a building, indicating the mobility 
is similar across user types. 
Figure \ref{fig:pri_sec_stat}(c) analyzes the  duration of each intra-building transition.
Such transitions result from users walking inside a building to see a colleague, go to a class or meeting, or to take a restroom break.  The CDF shows that the average transition within a building takes 1.5 minutes for faculty-staff and 1.48 minutes for students, which is again similar across user types.

Importantly, over the course of a day, the typical user makes 35.9 intra-building transitions across all visited buildings. Thus, we see 8X more mobility at intra-building (micro) scale than at inter-building (macro) scale, implying that mobility decreases at higher spatial scales (37.8 intra-building vs 4.1 inter-buildings locations visited).
Faculty-staff users make 14.8 intra-building transitions, while students make 37.9 transitions\footnote{Despite making fewer intra-building transition per visit, the higher number of buildings visited per day still yields an overall higher number of transitions for students.}; in doing
so, they spend 22.3 and 56 minutes walking inside buildings, respectively. Highly
mobile users representing the 90-th percentile of the distribution make 59.8 intra-building
transitions across all visited buildings, and spend a total of 90.4 minutes walking inside buildings.  


\begin{figure*}[ht]
\begin{minipage}[c]{0.48\linewidth}
\includegraphics[width=\textwidth]{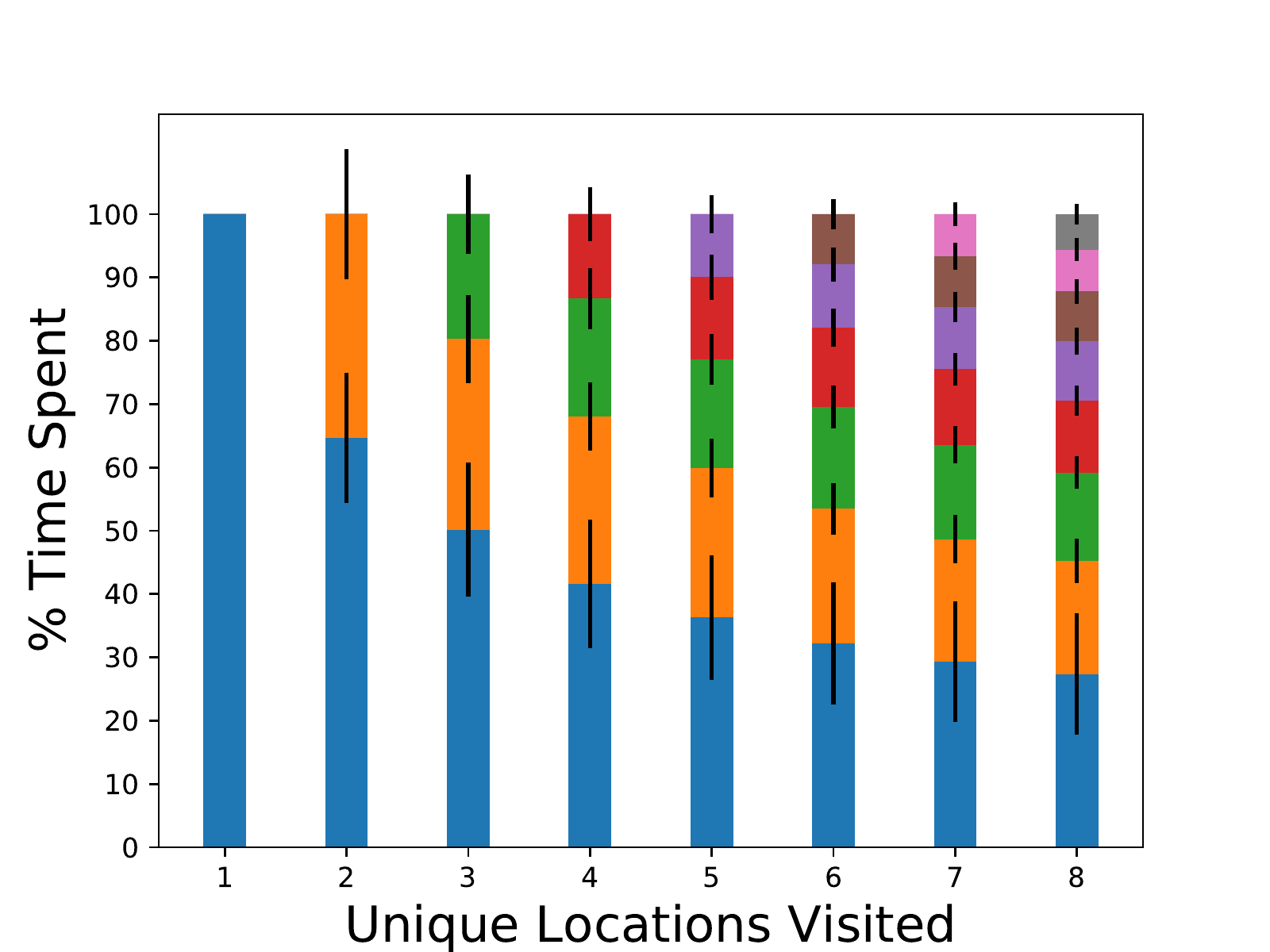}
\vspace{-0.2in}
\caption{Distribution of times spent across unique locations inside a building}
\vspace{-0.05in}
\label{fig:inside_bldg_time_spent}
\end{minipage}
\hfill
\begin{minipage}[c]{0.48\linewidth}
\includegraphics[width=\textwidth]{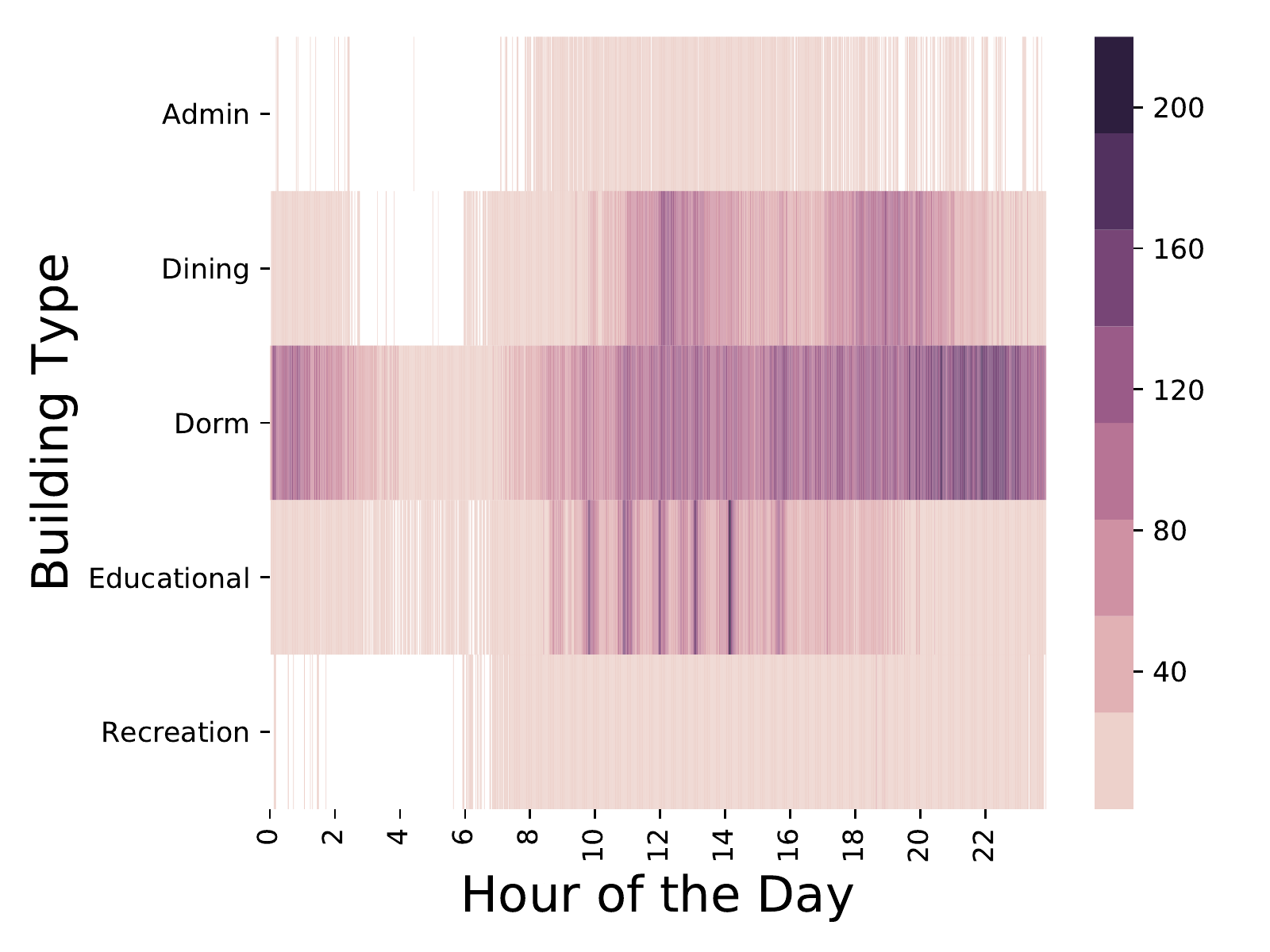}
\vspace{-0.2in}
\caption{Heatmap showing transitions by building type}
\vspace{-0.06in}
\label{fig:inter_bldg_heatmap}
\end{minipage}%
\end{figure*}

Figure \ref{fig:inside_bldg_time_spent} plots the distribution of time spent at  
unique locations visited inside a building versus the number of visited locations. 
Unlike the inter-building scale, where users spent 60 to 80\% of the time inside a single building, at inter-building scale, we find that users spend only 30 to 60\% of the time
at their most visited location; that number rises to  60\% or greater when we consider the top three most frequented location for each user. 


Finally, Figure \ref{fig:inter_bldg_heatmap} shows a heat map of when intra-building transitions occur over  the day. We find that mobility patterns inside a building are highly dependent on the type of the building---an academic building  sees very different intra-building mobility patterns than a residence hall. Figure \ref{fig:inter_bldg_heatmap} shows four different buildings from our overall analysis: a dining hall, a residence hall, an academic building, and an administration building. The figure shows residence hall users making intra-building transitions at all times of the day, while the academic buildings see transitions correlated with lecture start and end times as well as arrival and departure times. The dining halls see a high concentration of transitions at meal times such as breakfast, lunch, and dinner, while the administration building sees transitions during AM  arrivals and PM departures and uniform mobility in-between. More broadly, we find that the type of indoor space governs the type of intra-building mobility patterns that will be seen in that space.

\begin{mdframed}{
        \textit{ \textbf{Key Takeaway}}
    \begin{itemize}
            \item The typical campus user visits 8.97 locations inside each building during a building visit.  Over a day, a typical user visits 35.9 intra-building locations across all visited buildings.
            \item Users exhibit nearly 8X more mobility at intra-building scale than inter-building scale. 
            \item While the amount of mobility decreases (by 8X) with increasing spatial scale, the time spent at each visited location increases (by a factor of 2) with increasing scale. 
            \item Unlike inter-building mobility, users do not exhibit any affinity to a single intra-building location; over 60\% of the time during a building visit is spent at the three or fewer indoor locations.  
            \item The type of indoor space governs the  intra-building mobility patterns that are seen in that space.
        %
        \end{itemize} 
    }
\end{mdframed}

%% file: model.tex
\section{Mobility Modeling}
\label{sec:model}
Our results on multi-device users and mobility spatial scales ($\S$ \ref{multi_device_users} and \ref{dev_hierarchical_mobility})  have several implications on system design and modeling. We present two generative models, one each for modeling multi-device user and spatial scales, that leverage the insights from our analysis.

\subsection{Modeling Multi-device Users}  \label{find_my_device}

Our analysis in $\S$\ref{multi_device_users} showed that trajectories of a user's devices
are correlated and pointed to the need to model these dependencies, rather than treating them
as independent. To do so, we present a generative model that models the mobility of a group
of correlated devices. 

Our group model is based on a machine learning framework called \emph{multi task learning (MTL)} 
\cite{Caruana97} that is designed to learn multiple related tasks, while exploiting the similarities and differences between tasks. As noted in \cite{Caruana97}, doing so
can improve both learning and prediction efficiency, when compared to training models separately.
In our case, the task is one of predicting a device trajectory. We hypothesize that a MTL model
that models a collection of correlated devices will be richer and better than modeling each device independently.

\begin{figure}[h]
\centering
\includegraphics[height=1.5in]{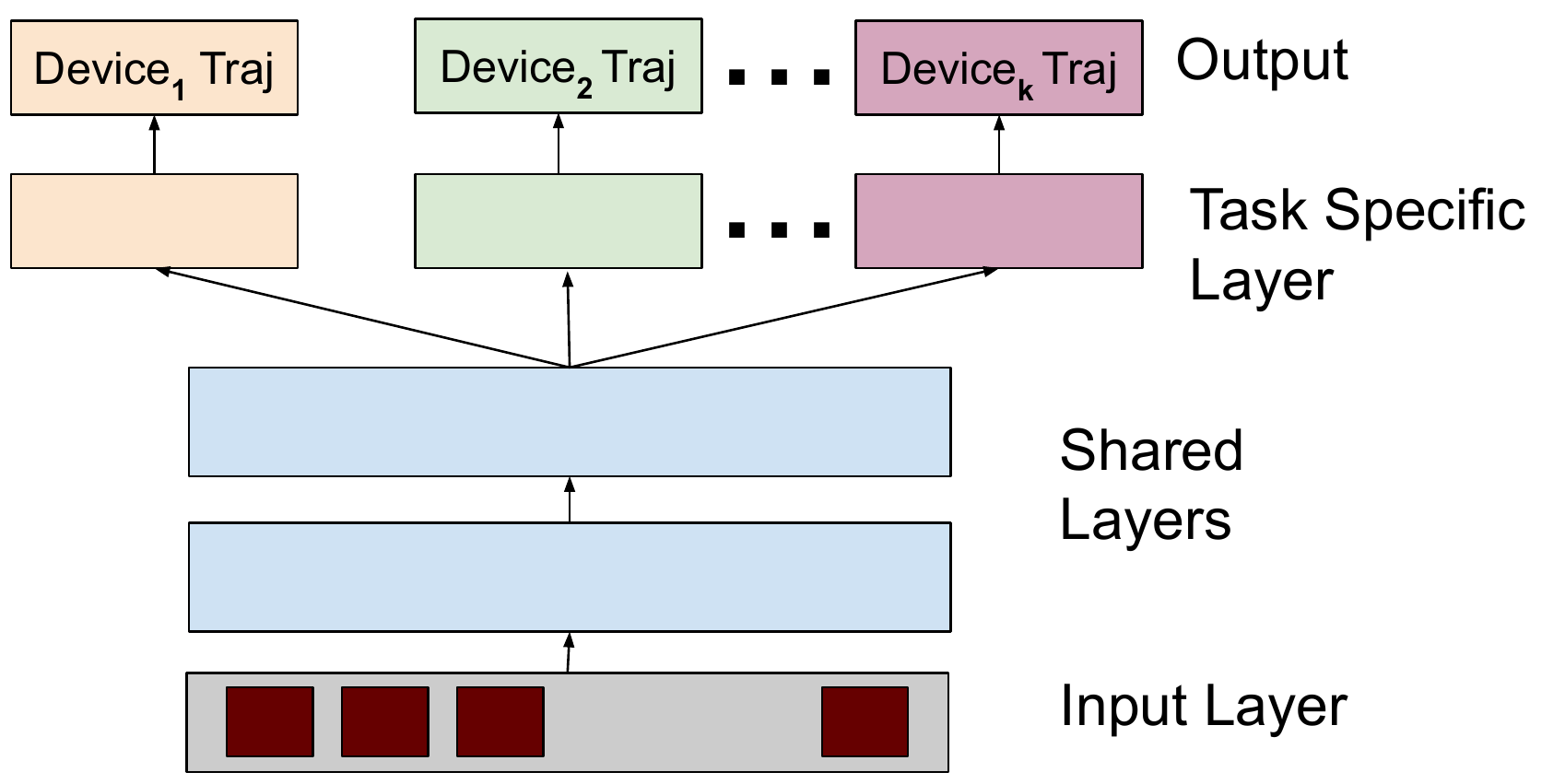}\\
\caption{MTL Architecture}
\vspace{-0.15in}
\label{fig:mlt}
\end{figure}

Figure \ref{fig:mlt} depicts the architecture of our MTL model. The model is initially trained
using trajectories of all $k$ devices belonging to a particular user. Once trained, 
the model takes the trajectory of a device within that group and predicts the
trajectory of the other $k-1$ devices for that period. This is feasible since the trained
model has learnt how the trajectories of devices within the group are correlated, and hence,
can use the input trajectory of a device to predict the rest.  As shown in \ref{fig:mlt}, 
we use a hard parameter sharing approach with two shared layers shared between all devices and one device-specific layer per device. Note also that since all $k$ devices share the same training data samples and same feature space, our MTL problem reduces to a homogenous feature multi-label learning problem. 

\begin{table*}[t]
\begin{minipage}[c]{0.3\linewidth}
     \begin{tabular}{lc} \toprule
     Input$\rightarrow$Ouput & Recall \\ \midrule
     Smartphone$\rightarrow$Tablet & 96.7\% \\
     Smartphone$\rightarrow$Laptop & 93.4\%\\
     Laptop$\rightarrow$Smartphone & 82.3\% \\
     Laptop$\rightarrow$Tablet & 81.9\% \\ \bottomrule
     \end{tabular}
     \caption{Prediction Efficacy}\label{tab:prediction_1}
\end{minipage}
\hfill
\begin{minipage}[c]{0.3\linewidth}
     \begin{tabular}{lc} \toprule
     Model & Accuracy \\ \midrule
     MTL & 67.32\% \\
     Individual Model & 35.24\% \\\bottomrule
    \end{tabular}
    \caption{Non-Conformance Prediction Efficacy(MTL v/s Individual Model)}\label{tab:prediction_nc}
\end{minipage}
\hfill
\begin{minipage}[c]{0.3\linewidth}
     \begin{tabular}{lc} \toprule
     Spatial Granularity & Accuracy \\ \midrule
     M1 (Inter-Building) & 92.96\% \\
     M2 (Intra-Building) & 57.49\%\\ \bottomrule
    \end{tabular}
     \caption{Prediction Accuracy}\label{tab:prediction_2}
\end{minipage}
\end{table*}

Such an MTL model that captures the mobility of a collection of related devices has many 
use cases. First, it allows a more compact method for capturing the mobility of each user. 
Once the model is trained for each use, we no longer need to track all of the user's devices.
Instead it is sufficient to track the mobility of a single device for that user since the model
can predict the mobility behavior of the remaining devices. In the future when each user
could own dozens of mobile devices, such a group model can be powerful method to track each
user's mobility in a compact manner using a single device while still capturing the behavior of the user's entire collection of devices.

Interestingly, it is not even necessary to use the user's most active device, i.e., their primary
device, to represent the user's mobile behavior. A MTL model is powerful enough to take 
a sparse trajectory of a (secondary) device such as a laptop and predict the dense trajectory of 
a phone, even though the laptop may not visit many of the locations visited by the phone.
MTL models can, of course, use the dense trajectory of a primary device to predict sparse
trajectories of secondary devices belonging to the user.

We trained MTL models for 100 different users and device collections from our dataset; the device groups represents varying degrees of correlations in mobility patterns discussed in 
$\S$\ref{multi_device_users}. We trained each model using data from 30 days. 
Once trained, we used each device belonging to a user to predict the trajectories of the remaining $k-1$ devices using the model, and compared the predictions to the ground-truth trajectory of those devices. 
As shown in table \ref{tab:prediction_1}, the average recall for predictions made from a sparse trajectory such as laptop to a dense trajectory such as a tablet or smartphone is 82.1\% while from a dense trajectory to a sparse trajectory is 95.1\%. Note that this prediction accuracy
is computed over a range of users with different degrees of correlations in their device trajectories.
Next, we compared our group model to predictions made by individual models learnt independently
for each device; each individual device model is based on a Long Short Term Memory(LSTM)-based RNN with RELU activation, MSE loss and adam optimizer.
An independent device LSTM model is well-suited for using trajectory from a previous day to predict one for the present day, but it can make erroneous prediction when the user's mobility
shows non-conformance from regularly observed repeating patterns. In contrast, an MTL model is robust to such non-conforming mobility behavior since it uses the current trajectory
of a device to predict those of the rest, and the non-conforming behavior is captured in the 
current trajectory used as model input. We carefully selected days when users had non-conforming
mobility patterns and used both the MTL and individual device models to predict trajctorries. 
As shown in Table \ref{tab:prediction_nc},  the MTL has atleast 32\% points higher accuracy during non-conformance behavior when compared to the predictions made by individual models. 
Together, these results confirm our hypothesis that group mobility modeling is superior to individual device mobility modeling due to the highly correlated device trajectories.

\subsection{Generative Model at Multiple Spatial Scales}
\label{next_loc}

\begin{figure}
\centering
\includegraphics[height=1.2in]{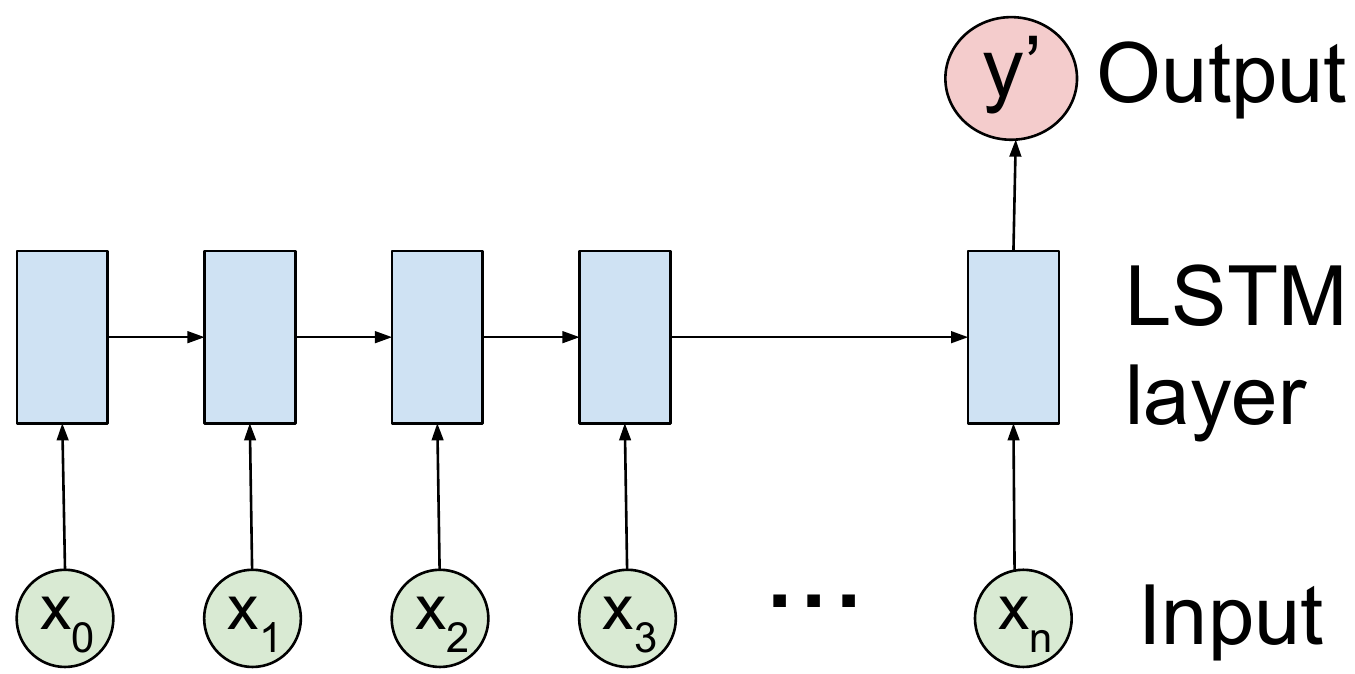}\\
\caption{LSTM Architecture}
\label{fig:lstm}
\end{figure}

Our results in $\S$\ref{dev_hierarchical_mobility} highlighted the importance of using an
appropriate spatial scale when addressing mobility problems. In this section, we show 
that choice of the ``correct'' spatial scale can make a significant difference 
when addressing common mobility problems. To do so, we consider the next location prediction
problem, which is a widely studied problem in mobile computing \cite{Ganti:2013:IHM:2493432.2493466,Do:2012:CCM:2370216.2370242,liu2016predicting,song2004evaluating}.
Consider a location-aware mobile service that uses next location prediction in a campus
setting to predict the next building visited by a user in order to prefetch useful
information such as a list of current events and a building map before the user arrives at the building. Given our WiFi dataset, if device trajectories are modeled at the spatial
scale of AP-level locations (i.e., intra-building scale), a next location prediction model
will predict the next AP location visited by the user, which may be a location in the current
building. In fact, a sequence of next $k$ AP locations need to be predicted by the model
to determine the next building. As noted in $\S$\ref{dev_hierarchical_mobility}, a user
visits 9 locations, on average, inside each building, implying that a sequence of 9 next locations needs to be predicted to determine the next building, on average. 
On the other hand, if we model device trajectories at  inter-building spatial scale, the problem
of determining the next building becomes a straightforward task of predicting the next location. 
Clearly, the accuracy of predicting the immediate next location will be higher than a sequence
of 9 future locations. 

To demonstrate this difference, we use a LSTM-based RNN model to perform next location predictions both at inter-building and intra-building
spatial scales. We use the LSTM model to predict a sequence of future locations.
Our inter-building LSTM model is trained on past trajectories of building visits made by
a group of users, while the intra-building LSTM model is trained on the same trajectories but using AP-level location visits. Both LSTM models have RELU activation, MSE loss and adam optimizer.. Figure \ref{fig:lstm} depicts these models.
We use the inter-building LSTM model to predict the  next building visit, given the immediate 
past trajectory of a user. For the intra-building LSTM model, we predict a sequence of 
9 next APs visited and use this sequence to determine the next building location. 

\begin{figure}[h]
\begin{tabular}{cc}
\includegraphics[width=0.45\linewidth]{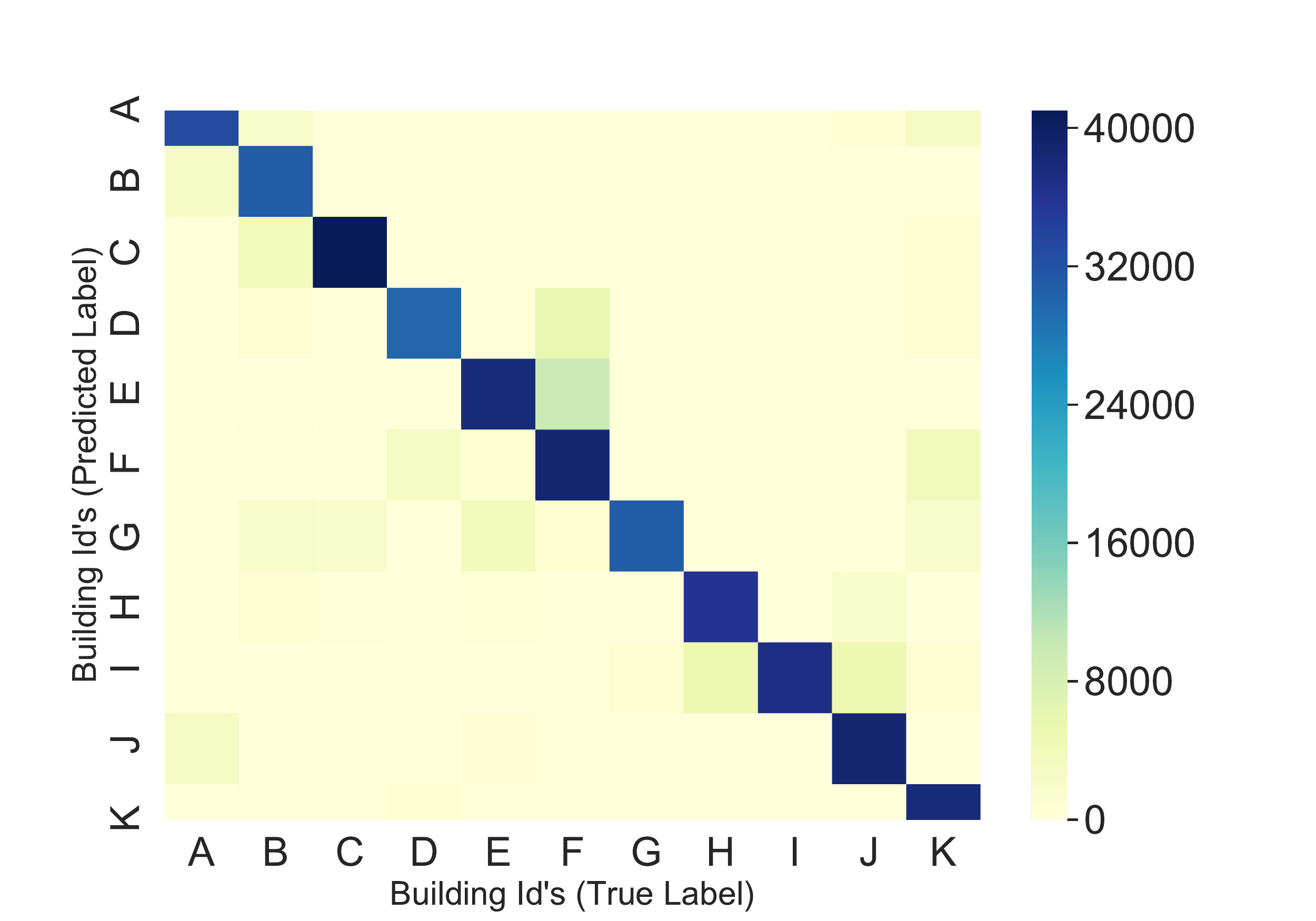}&
\includegraphics[width=0.45\linewidth]{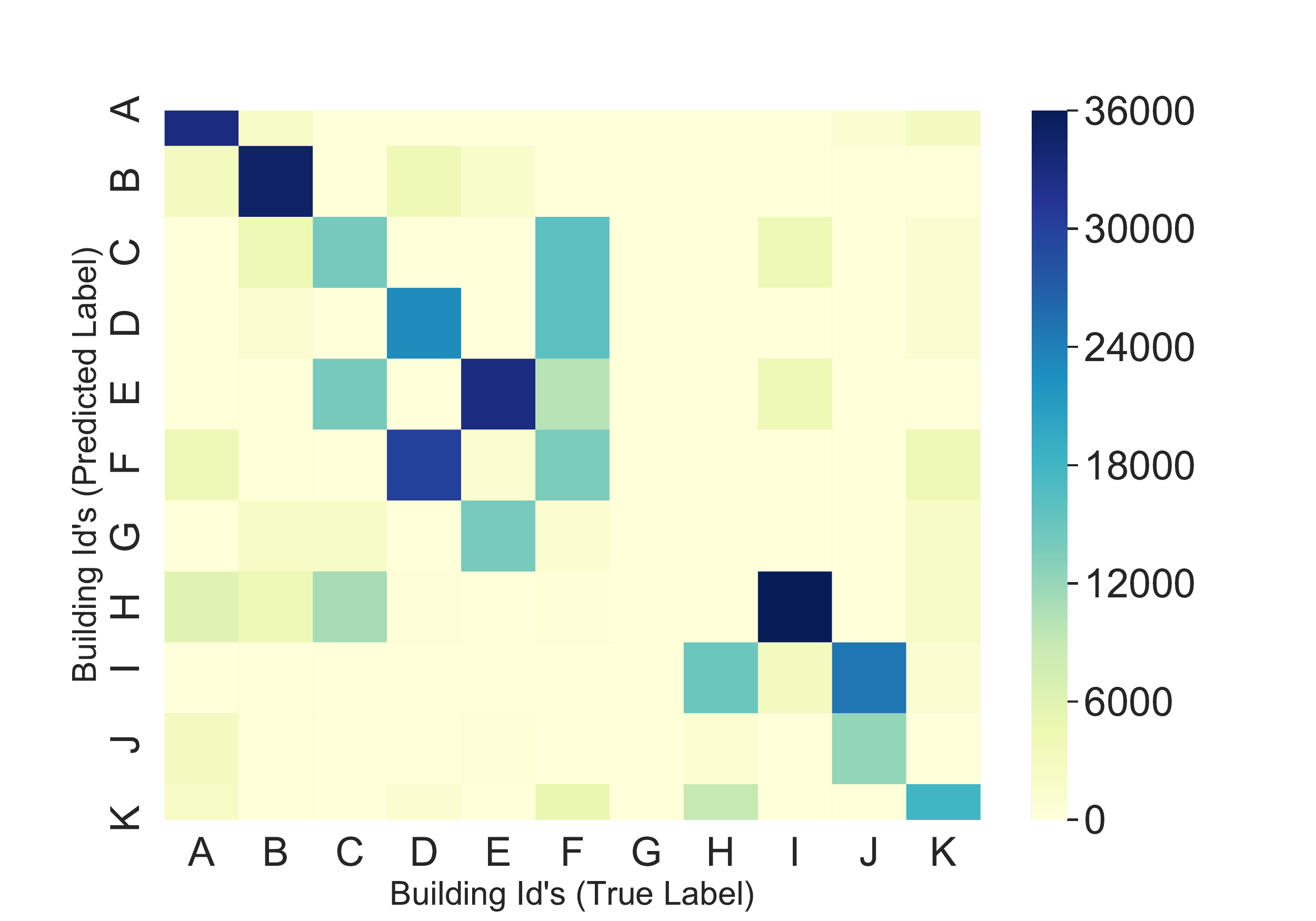}\\
(a) & (b) \\
\end{tabular}
\caption{Confusion matrix for (a) inter-building and (b) intra-building LSTM models.}
\label{fig:M1_confusion}
\end{figure}

As shown in Table \ref{tab:prediction_2}, we find that the inter-building model has an mean accuracy of 92.9\%, while the intra-building model has a much  lower accuracy of  57.5\% when trained on the same dataset containing 1500 users.
Figure \ref{fig:M1_confusion}(a) and (b)  show the confusion matrix of predictions made by inter- and intra-building LSTM models, with the inter-building spatial scale model yielding better accuracy. Together, these results highlight how choice of the spatial scale can enhance the accuracy of a mobility model (here, LSTMs) when performing tasks such as next location predictions.

%% file: implications.tex
\section{Implications of Our Results}
\label{sec:implications}

We now discuss the broader implications of our results.

{\bf Campus-scale Mobility:}
Overall, we find that campus-scale mobility in building depends on {\em five} key factors:

\emph{Spatial Scale:} Our results show that as the spatial scale becomes coarser, the amount of  mobility in terms of locations visited decreases with a corresponding
increase in time spent at each visited location. We found that intra-building mobility
was 8X more frequent with shorter stays at each location than inter-building mobility. 
Since  mobility is more frequent at micro scales than macro scales, a judicious choice of the correct spatial scale is necessary when addressing system design problems, as shown in $\S$\ref{sec:model}.

\emph{Device type:} Our results indicate that less portable devices have lower mobility, 
since phones were found to be 3.5X more mobile than laptops.
This key finding implies that all mobile devices should not be treated as equal and 
optimizing systems based on device type (or size-based groupings) may yield a better overall  design.

\emph{Multi-device ownership:} Given the prevalence of multi-device ownership, treating devices as being independent of others is no longer a reasonable approach. Our results showed
that device trajectories of various devices owned by a user exhibit moderate to strong 
correlations but also that the degree of correlations varies considerably from user to user. Thus jointly modeling group of devices owned
by a user or exploiting mobility pattern of one device to predict those of others for that user may yield better results, as shown in $\S$\ref{sec:model}.

\emph{User behavior:} Some users will naturally be more mobile than others, and this mobile behavior manifests differently at different scales. At a given scale, highly-mobile 
users visit several times more locations than the average user. Across spatial scales, users who visit more buildings per day are {\em less mobile}, on a per-visit basis, at the intra-building scale, since their higher inter-building mobility results in shorter stays and few intra-building location visits per building.  These findings manifested themselves in our study as students being more mobile, as a group, at inter-building scales,
and faculty-staff, as a group, visiting more locations per building visit at intra-building scale.

\emph{Building type:} Our results show that the intra-building mobility patterns are heavily dependent on the type of the building; the functions served by a building determine how frequently and when indoor mobility will be seen.  The same user will exhibit different mobile behavior in different types of buildings, which implies that mobile behavior is not just a user characteristic but also depends on the building type.

{\bf Outdoor versus Indoor Mobility:} Our study reveals important differences between
outdoor and indoor mobility and also some similarities. First, similar to the findings in \cite{zhou2017mining}, we find that mobility in buildings  is far more frequent than urban-scale outdoor mobility in terms of the number of locations visited. Of course, transition times and distance traveled will naturally be smaller inside buildings than in outdoor spaces. 
Thus, results from outdoor mobility should not be directly employed when designing systems
that will be primarily deployed and used inside buildings or on campuses. Interestingly, outdoor mobility can be viewed as a natural progression in the hierarchy from inter-building mobility, and when viewed from this standpoint, it naturally follows that mobility (in term of number of locations visited) will be lower at outdoor scales than finer indoor building scales---in line with our hierarchical spatial scale findings.  A hierarchical study that combines both outdoor and indoor mobility in an integrated fashion is left to future work. 

%% file: conclusion.tex
\section{Conclusions and Future Work}
\label{sec:conclusion}

This paper presented an analysis and modeling of mobility of multi-device users
in a university campus. Our study showed that mobility decreases with spatial scale---with 8X less mobility at inter-building scale than intra-building scale. We also found that  mobility is  related to device type---phones have 3.5X greater mobility than laptops. Despite these differences, devices belonging to the same user show moderate to strong correlations in mobility for the majority of the users, and the type of building has a significant influence on the frequency and timing of mobility patterns observed in it. Our generative models
highlighted how these insights can be exploited to build better models to capture campus-scale
user mobility patterns.
Overall, our study revealed significant differences between mobility in buildings and prior results on urban-scale outdoor mobility, indicating that system designers need to carefully consider our findings when designing systems for indoor use. 


%% file: appendix.tex
\appendix

\section{Smoothing and Filtering over Noisy Trajectories}
\label{sec:smoothing}
To generate a clean trajectory trace from the inherently noisy raw WiFi logs, we use a combination of smoothing and filtering techniques. We use two techniques, one for removing 
noise during stationary periods and another to deal with noise during walking transitions. 
For stationary periods, the main source of noise occurs from devices that ping-pong
between nearby APs when being stationary. To address this noise, we consider the AP with
the longest association duration during a stationary period as the true location and filter brief associations with other nearby APs as noise in the midst of a longer stationary period.  It is also possible that a device may continuously connect to a more
distant AP than the closest one; if the device remains continuously connected to a 
AP that is a little farther away than the nearest AP to the true location, we simply treat this as measurement error rather than noise. Such effects can be seen in densely instrumented areas such as an auditorium with multiple nearby APs and there is no guarantee that the device connects to the closest one; however the measurement error in estimated location is usually small in such densely instrumented areas. Our analysis of 
the data as well as our validation study do not show devices choosing to {\em continuously} connect to very far-off APs over more proximal ones with better signal; for example, our validation data did not observe any devices continuously connecting to APs on a different floor than the true location especially since all campus buildings are well instrumented with nearby APs with good signal quality; of course if such more distant APs are briefly, rather than continuously, chosen by a device due to ping-ping effects, the above filtering
technique will eliminate such noise.  Finally, we use a distance-based smoothing heuristic
to handle noise during walking transitions.  Walking transitions involve an always-on
device choosing to associate with a changing set of APs along the walking trajectory. 
We assume that the human walking speed  is limited (e.g., less than 2meters/s) and locations cannot change by a distance that exceeds a threshold walking speed. Thus, if a device associates with a distant building AP during an outdoor walk and then again with a nearby one, such a technique will smoothen the trace by eliminating distant "jumps" in locations during a walk. 
Such a heuristic is useful during outdoor walks between buildings since outdoor WiFi coverage is weaker than indoors;  indoor walks can also use such heuristics but they
are less likely to see such effects due to the higher density of indoor deployments, which allows devices to choose proximal APs rather than very distant ones.

\section{Trajectory Validation}
\label{sec:validate}

 \begin{figure}[ht]
\center
\includegraphics[height=2.4in]{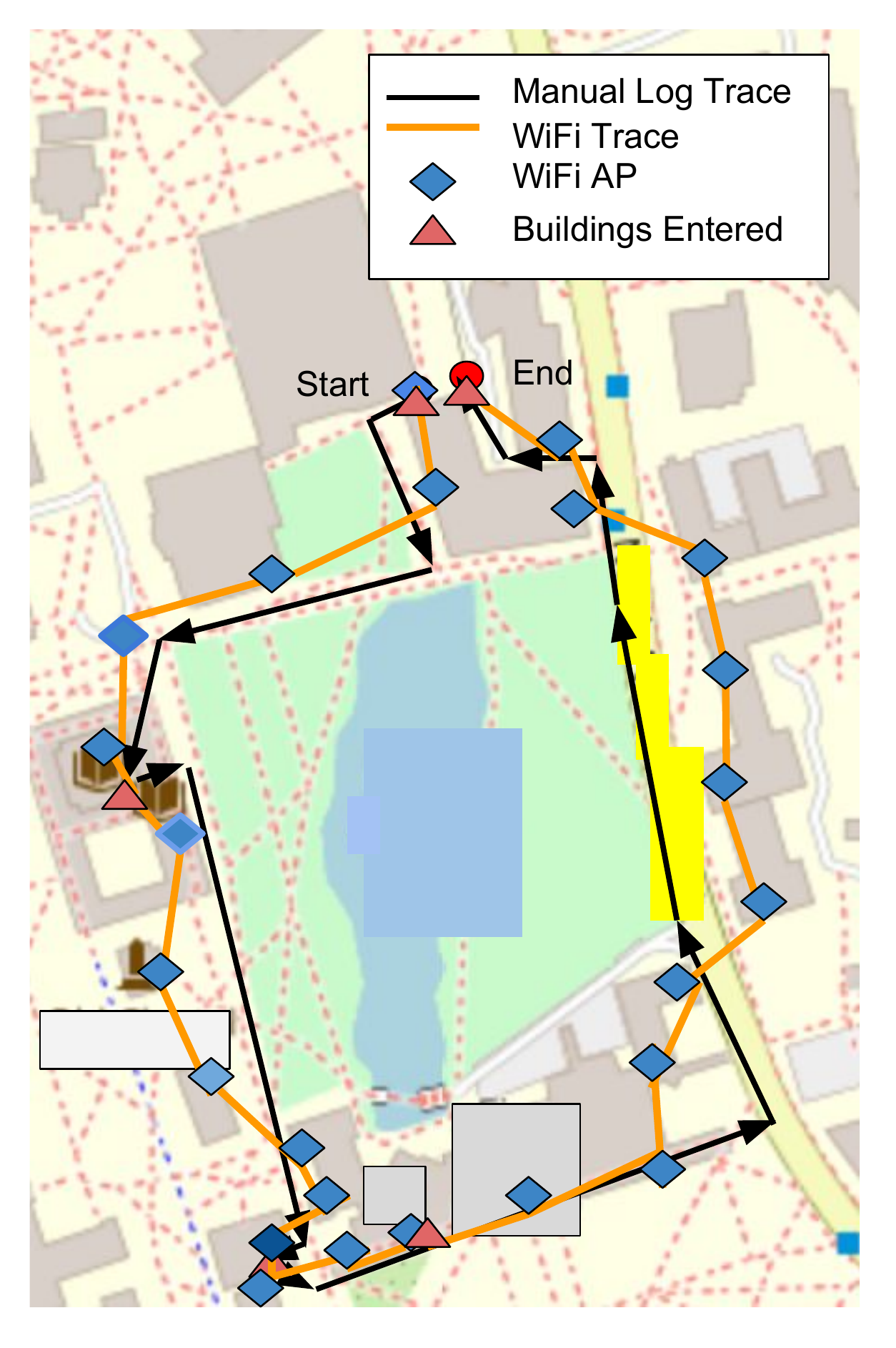}
\caption{Comparison of WiFi trajectory of a device across campus with the ground truth.}
\vspace{-0.05in}
\label{fig:validation1_plots}
\end{figure}

 The use of location data from mobile devices has previously been used for  analyzing mobility behavior \cite{kim2006extracting,hang2018exploring} and is considered to be a valid method for such studies.
Nevertheless we conduct a  small-scale  validation study of our WiFi dataset to verify its
efficacy for extracting device trajectory. First, to validate that device trajectories derived from the WiFi dataset corresponds to the actual device trajectory, 
we had volunteers mimic user behavior by walking with a phone to various campus buildings, spend some time inside each building, and then walk to next building and so on. The ground truth trajectory (i.e., locations and times) recorded by the user were compared to that derived from the WiFi log of the phone. Figure \ref{fig:validation1_plots} depicts
one sample trajectory from our validation experiments. As shown, the computed trajectory closely matches the ground truth, indicating that campus WiFi coverage is sufficient to 
track device movements inside buildings and even outside buildings. For all indoor locations and stationary periods, we observed near perfect accuracy between the ground truth and observed locations. We did observe some deviations between ground truth and 
estimated location during outdoor walks between buildings due to weaker outdoor WiFi coverage (see Figure \ref{fig:validation1_plots});  in these scenarios, the deviation was around 20-40 meters (typically the distance to the closest building to the walking path). Given our focus on indoor mobility, such deviations in location estimates in outdoor settings do not impact the majority of indoor analysis results.

\begin{figure}[t]
\includegraphics[width=2.5in]{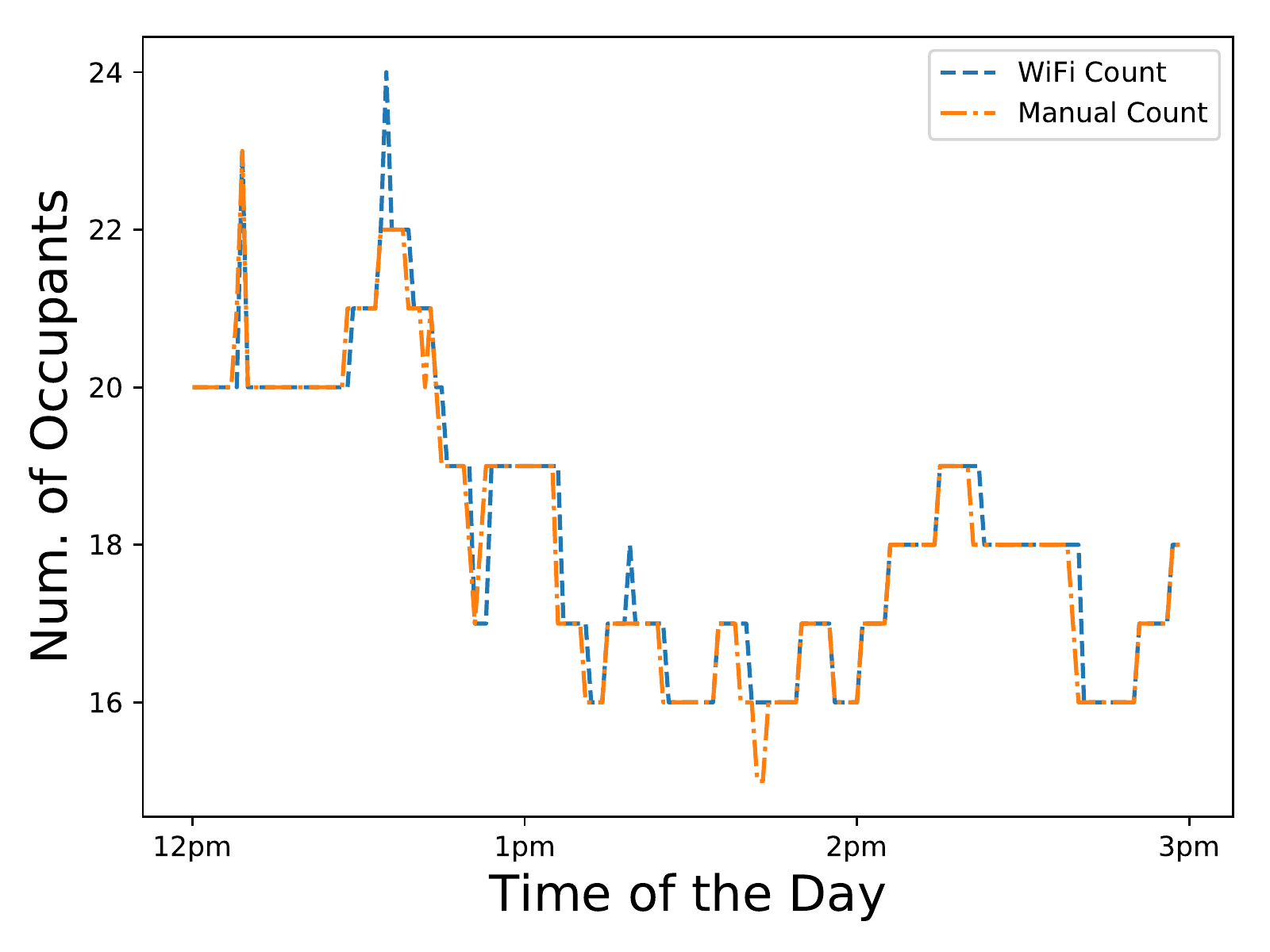}
\caption{Comparison of users visible to the WiFi network at a busy campus location and the ground truth.}
\vspace{-0.05in}
\label{fig:validation2_plots}
\end{figure}

Second, we asked a  volunteer to  visit several popular locations on campus that see substantial user traffic such as the  campus center, academic buildings, and dining halls and counted the number of users entering and leaving popular areas. This ground truth data was compared to the number of users reported by access points  in the vicinity of each location. Figure \ref{fig:validation2_plots} depicts the number of users visible to the WiFi network over time at one such location and the ground truth. As shown, the two values match closely, indicating that most users carry at least one mobile device with them while on campus, and our WiFi dataset is therefore able to provide near-comprehensive coverage of users on this campus for our mobility study.


